\documentclass[twocolumn,prb,showpacs,floatfix]{revtex4}
\usepackage{graphicx}
\usepackage{dcolumn}
\usepackage{amsmath,amssymb} 
\usepackage{amsmath,amssymb}

\newcommand{\beq}{\begin{equation}}
\newcommand{\eeq}{\end{equation}}
\newcommand{\beqa}{\begin{eqnarray}}
\newcommand{\eeqa}{\end{eqnarray}}

\newcommand{\D}{\Delta}
\newcommand{\e}{\epsilon}
\newcommand{\sx}{\sigma_x}
\newcommand{\sz}{\sigma_z}
\newcommand{\sy}{\sigma_y}
\newcommand{\s}{\alpha}		
\newcommand{\sprime}{\beta}		

\renewcommand{\l}{\eta}		
\renewcommand{\a}{\alpha}	
\renewcommand{\t}{\tau}
\renewcommand{\b}{\beta}
\renewcommand{\o}{\omega}
\newcommand{\ad}{\gamma}	

\newcommand{\cl}{\ell}

\newcommand{\kvec}{{k}}	

\begin{document}
\title{Polaron Crossover and Bipolaronic Metal-Insulator Transition in the
Holstein model at half-filling.}

\author{M. Capone}

\affiliation{INFM-SMC and Istituto dei Sistemi Complessi,
Consiglio Nazionale delle Ricerche,
via dei Taurini 19, I-00185 Rome, Italy}

\author{P. Carta}

\affiliation{Universit\`a degli Studi di Cagliari, Dipartimento di Fisica and
INFN, Sezione di Cagliari, Cittadella Universitaria I-09042, Monserrato, Italy}

\altaffiliation{
Present address: The Royal Bank of Scotland.
Financial Markets
135 Bishopsgate EC2M 3UR London, UK}

\author{S. Ciuchi}

\affiliation{Dipartimento di Fisica and INFM, Universit\`{a} de L'Aquila,
67010 Coppito-L'Aquila, Italy}


\date{\today}
\begin{abstract}

The evolution of the properties of a finite density electronic system
as the electron-phonon coupling is increased are investigated in the
 Holstein model using the Dynamical Mean-Field Theory (DMFT).

We compare the spinless
fermion case, in which only isolated polarons can be formed,
with the spinful model
in which the polarons can bind and form bipolarons. In the latter case, the
bipolaronic binding occurs through a  metal-insulator transition.
In the adiabatic regime in which the phonon energy is small with respect to
the electron hopping we compare numerically exact DMFT results with
an analytical scheme inspired by the Born-Oppenheimer procedure.
Within the latter approach,a truncation of the phononic
Hilbert space leads to a mapping of the original model onto an Anderson
spin-fermion model. In the anti-adiabatic regime (where the phonon energy
exceeds the electronic scales) the standard treatment based on Lang-Firsov
canonical transformation allows to map the original model on to an attractive
Hubbard model in the spinful case. The separate analysis of the two regimes
supports the numerical evidence that polaron formation is not necessarily
associated to a metal-insulator transition, which is instead due to pairing
between the carriers.
At the polaron crossover the Born-Oppenheimer approximation is shown to break
down due to the entanglement of the electron-phonon state.

\end{abstract}
\maketitle

\section{Introduction}

Electron-phonon (e-ph) interaction play an important role in virtually all the
materials of present interest. To be concrete, e-ph coupling is most likely the
driving force for superconductivity in  magnesium diboride\cite{mgb2}, in the
alkali-doped fullerides\cite{revgunnarson} and in recently studied
intercalated graphite compounds\cite{graphite}.
Despite the central role of electron-electron
correlation, also the high-$T_c$ cuprates are now believed to display
remarkable e-ph features\cite{cuprates-lanzara-dastuto}
witnessed by isotope effects\cite{lanzara-isot} as well as spectral and
transport properties \cite{calvani-rev}.
Jahn-Teller e-ph interaction is one of the key interactions
in the colossal magnetoresistance manganites\cite{manganites,IR-Manga},
and it may be important in
transition-metal oxides\cite{oxides}.
Last, but not least, many different families
of organic materials are characterized by coupling with ionic
degrees of freedom, from nanotubes to DNA\cite{organics}.
Given the broad variety of different
physical properties and origin of the coupling, such a wealth of materials
covers basically all the different regimes characteristic of the
e-ph interaction.
In manganites and cuprates as well as in most of the oxides-based materials,
the electronic bandwidths, even if renormalized by strong e-e interactions, are
larger than phonon frequencies leading to an adiabatic character of e-ph
interaction, in which the Born-Oppenheimer framework is, at least, the
reference framework.
This is not the case of highly-polarizable crystal lattices of
organic semiconductors \cite{Morpurgo}. In this materials molecules are bound
by very weak Van der Waals forces while intramolecular phonons can have quite
large oscillation frequencies leading to an intrinsic anti-adiabatic regime of
e-ph interaction, in which the adiabatic principle breaks down.

In system with strong e-ph coupling, the carriers lose mobility, eventually
acquiring polaronic character. A polaron is a state in which the phonon and
electron degrees of freedom are strongly entangled, and the presence
of an electron is associated to a finite lattice distortion, which in
turn binds the electron leading to the so-called self-trapping effect.
Polarons also tend to create bound pairs, called bipolarons.
One of the purposes of our studies is to clearly distinguish between
polaronic and bipolaronic features which are often confused  in
literature.

The aim of this work is to provide a thorough analysis of the Holstein model
at half-filing, comparing spinless and spinful fermions and discussing in
detail the role of the adiabatic ratio. The backbone of our presentation is a
numerically exact solution of the Dynamical Mean-Field Theory (DMFT)
a quantum version of mean-field approaches which does not rely on any small
parameter assumption. DMFT has been already applied to the study of polaronic
systems using analytical result for the impurity solution at low density
\cite{depolarone,frat2}.
At finite density exact  analytical method are not available and we
use Exact Diagonalization (ED) as an impurity solver.
Such results allow us to identify the virtues and defects of various analytic
approximate schemes. In particular we discuss in detail a Born-Oppenheimer (BO)
scheme which is based on the adiabatic limit,  and discuss an antiadiabatic
approach slightly different from the popular Lang-Firsov based  approaches to
polaronic systems \cite{Lang-Firsov}. In particular the BO scheme which will be
presented here is suitable to be applied to {\it both} the weak and the strong
coupling regime.

We organize our presentation as follows:
After a brief introduction of the model and of its treatment
within DMFT (Sec. \ref{sec:model}),
we anticipate the main results from the exact
DMFT comparing numerical results for different regimes (Sec. \ref{sec:results}).
Then we discuss in detail the Born-Oppenheimer approach and compare it with
numerical results in the adiabatic regime (Sec. \ref{sec:adiabatic}), and
analogously we compare the Lang-Firsov approach with results in the antiadiabatic
regime in Sec. \ref{sec:antiadiabatic}). Concluding remarks are reported in Sec.
\ref{sec:conclusions}. In appendix are reported the details of the calculations
used in Sec. \ref{sec:adiabatic}.

\section{The model in DMFT scheme}
\label{sec:model}

The Holstein Hamiltonian is perhaps the simplest lattice model to describe
e-ph interactions. Tight-binding electrons are coupled
to dispersionless local vibrational modes. The Hamiltonian is
\begin{eqnarray}
\label{eq:themodel}
H &=& -t\sum_{\langle i,j\rangle,\sigma} (c^{\dagger}_{i,\sigma} c_{j,\sigma} +
h.c. ) - g\sum_{i,\sigma} (n_{i,\sigma}-\frac{1}{2})(a_i +a^{\dagger}_i) +
\nonumber\\
&+& \omega_0 \sum_i a^{\dagger}_i a_i,
\end{eqnarray}
where $c_{i,\sigma}$ ($c^{\dagger}_{i,\sigma}$) and $a_i$ ($a^{\dagger}_i$) are, respectively,
 destruction (creation) operators for
fermions and for local vibrations of frequency $\omega_0$
on site $i$, $n_{i,\sigma}=c^{\dagger}_{i,\sigma}c_{i,\sigma}$
the electron density per spin, $t$ is
the hopping amplitude, $g$ is an electron phonon coupling.
We always fix the chemical potential to the particle-hole
symmetric value, which fixes the density per spin to $n = 1/2$.
In the spinless case there is no sum on $\sigma$.
We choose as parameter of the model the electron-phonon coupling constant
$\lambda = 2g^2/\omega_0 D$ where $D$ is the half-bandwidth of the electrons,
and the adiabatic ratio $\ad = \omega_0/D$.

In DMFT, the lattice model is mapped onto an impurity
problem subject to a self-consistency condition, which contains
all the information about the original lattice.
Our model (\ref{eq:themodel}) becomes  a Holstein impurity model (HIM),
\begin{eqnarray}
\label{eq:Anderson_Holstein}
H &=& -\sum_{k,\sigma} V_k (c^{\dagger}_{k,\sigma} f_{\sigma} + h.c)
\sum_{k,\sigma} E_k c^{\dagger}_{k,\sigma} c_{k,\sigma} \nonumber\\
&-&
g \left (\sum_\sigma f^\dagger_\sigma f_\sigma - \frac{1}{2} \right)(a +a^{\dagger}) + \omega_0 a^{\dagger}  a,
\end{eqnarray}
where the phonons are defined only on the impurity site 0, and they interact with
the electrons that jump on that site.
For the Bethe lattice of half-bandwidth $D$
the self-consistency enforcing the DMFT solution is given by \cite{DMFTreview}
\begin{equation}
\label{eq:self-cons}
\frac{D^2}{4}G(i\omega_n) = \sum_k \frac{V_k^2}{i\omega_n - E_k}.
\end{equation}
where $G(i\omega_n)$ is the local Green's function of the system. One could
eventually get the electron self-energy trough the following relation which
holds in the Bethe lattice case
\begin{equation}
\label{eq:sigma}
G(\omega) =\frac{1}{\omega-\frac{D^2}{4}G(\omega)-\Sigma(\omega)}.
\end{equation}
  
We solve the HIM by means of exact diagonalization, i.e., by truncating the sums
in the first two terms of Eq. (\ref{eq:Anderson_Holstein}) to a small number
of terms $N_b$, so that the Hilbert space is small enough to use, e.g., the
Lanczos algorithm to compute the $T=0$ Green's function.
For the case of phonon degrees of freedom we consider here, also the infinite
phonon space has to be truncated allowing for a maximum number of excited
phonons $N_{ph}$. In all the calculations presented here the convergence of
both truncations have been checked. The value of $N_{ph}$ has to
be chosen with special care in the adiabatic regime and in strong coupling,
where phonon excitations are energetically convenient.
As far as the discretization of the bath is concerned, the convergence of
thermodynamic averages and Matsubara frequency properties is exponentially
fast and $N_b \sim 8-9$ is enough to obtain converged results.
The method also offers the advantage of a direct evaluation of real-frequency
spectral properties such as the electron and phonon spectral functions.
The main limitation is that these quantities reflect the discrete nature of
our system much more than their imaginary-frequency counterparts. In practice,
the spectra are formed by collections of $\delta$-functions. Of course this
limits our frequency resolution, and suggests that the method is better
suited to gain knowledge on the main features of the spectra, rather than
on the fine details.

In the following we define some important quantities which we use to
discuss and characterize the physics of our model. Our focus is mainly on
the polaron crossover and the metal-insulator transition.
The quasiparticle residue $Z = (1-\partial\Sigma(\omega)/\partial\omega)\vert_{\omega=0}^{-1}$,
$\Sigma(\omega)$ being the electron self-energy, proved extremely useful
in marking metal-insulator transitions, both for repulsive and attractive
models, and it has already been employed for the characterization
of the bipolaron transition \cite{Max1}.
More detailed informations on the passage from a metal to an
insulator can be gained through the study  of the electron spectral function
(or density of states)
\begin{equation}
\label{eq:defDOS}
\rho(\omega)=-\frac{1}{\pi} Im G(\omega).
\end{equation}

As far as the polaron crossover is concerned, a key quantity is
the lattice displacements probability distribution function (PDF)
\cite{Millis-adiab,cdw-adiab}
\begin{equation}
\label{eq:defPDF}
P(X)=\langle\psi_0\vert X\rangle \langle X\vert\psi_0\rangle,
\end{equation}
where $\vert\psi_0\rangle$ is the groundstate vector of the impurity
model and $X$ is the phonon displacement operator on the impurity site.
The appearance of a bimodal PDF signals indeed at finite densities
the formation of a polaronic state, i.e., a state in which the presence
of the electron is associated to a definite polarization of the lattice.
At finite densities and for strong coupling the electrons are able to
drastically change the phonon properties, as opposed to the case of a single
polaron\cite{Millis-adiab}.
In the spinful case such a state
may be bipolaronic, i.e., the polarization of the lattice may be associated to a
local pairing of a pair of electrons with opposite spin.

Further information on the polaronic properties and on the mutual interaction
between electrons and phonons can be extracted from the phonon propagator
\begin{equation}
\label{eq:defD}
D(t)= -i \left \langle T (a^\dagger(t) + a(t))(a^\dagger(0) + a(0)) \right \rangle,
\end{equation}
from which the phonon spectral function is readily defined as
\begin{equation}
\label{eq:defDOSPH}
\rho_{ph}(\omega)=-\frac{1}{\pi} Im D(\omega).
\end{equation}

\section{Results}
\label{sec:results}
\begin{figure*}[htbp]
\begin{center}
\includegraphics[scale=0.7]{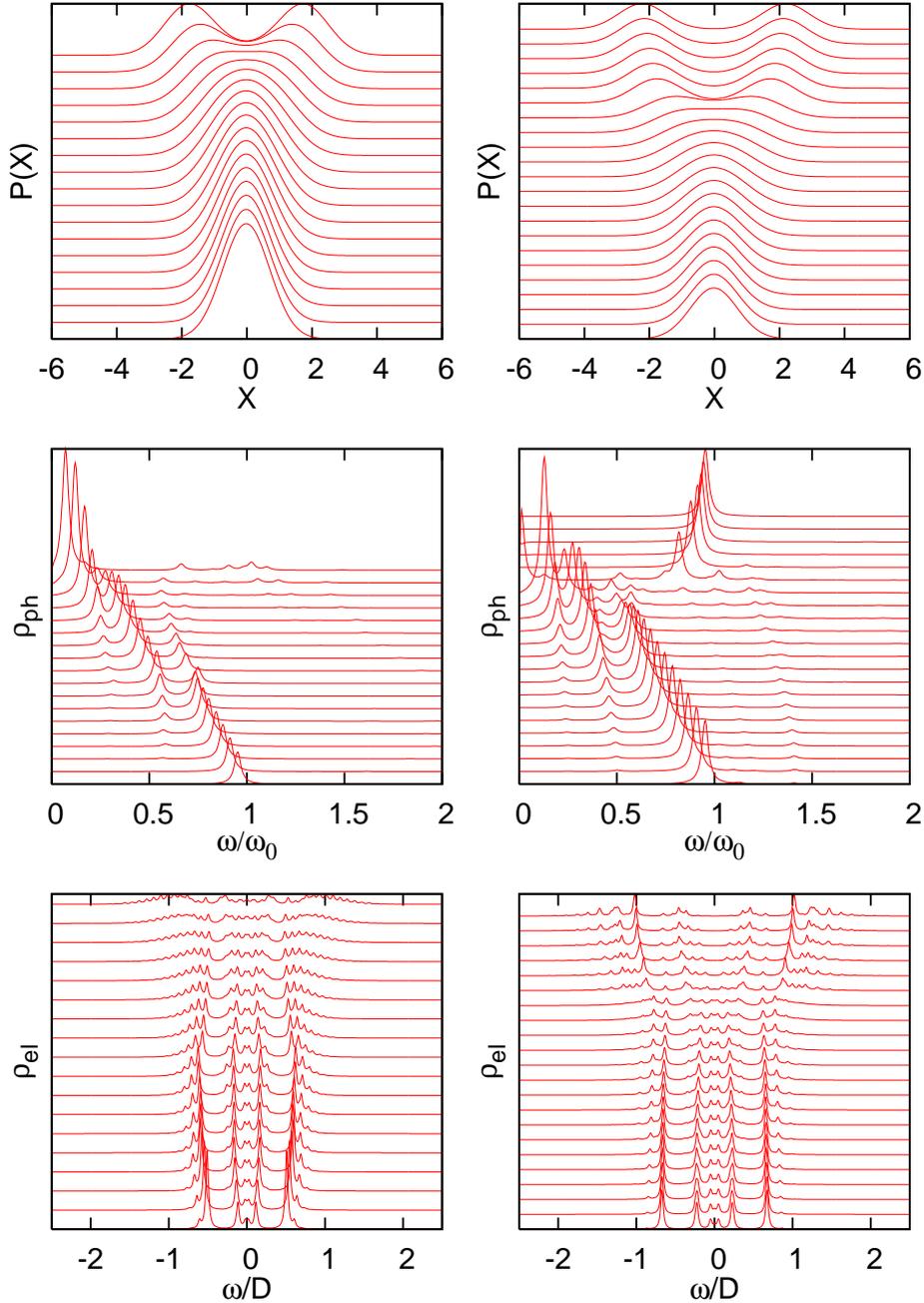}
\end{center}
\caption{(color online) DMFT data in the adiabatic regime $\ad=0.1$ spinless
(panels on the left) and
$\ad=0.2$ spinful (panels on the right). In each panel the various Curves 
refer to different value of
$\lambda$ spanning from  $0.1$ to $1.8$ in the spinless case
and from $0.05$ to $1.1$  in the spinful case
and are shifted according $\lambda$ value.
The first line show the phonon PDF, the central line the phonon DOS
and bottom line the  electronic DOS.}
\label{fig:data_adiab}
\end{figure*}

\begin{figure*}[htbp]
\begin{center}
\includegraphics[scale=0.7]{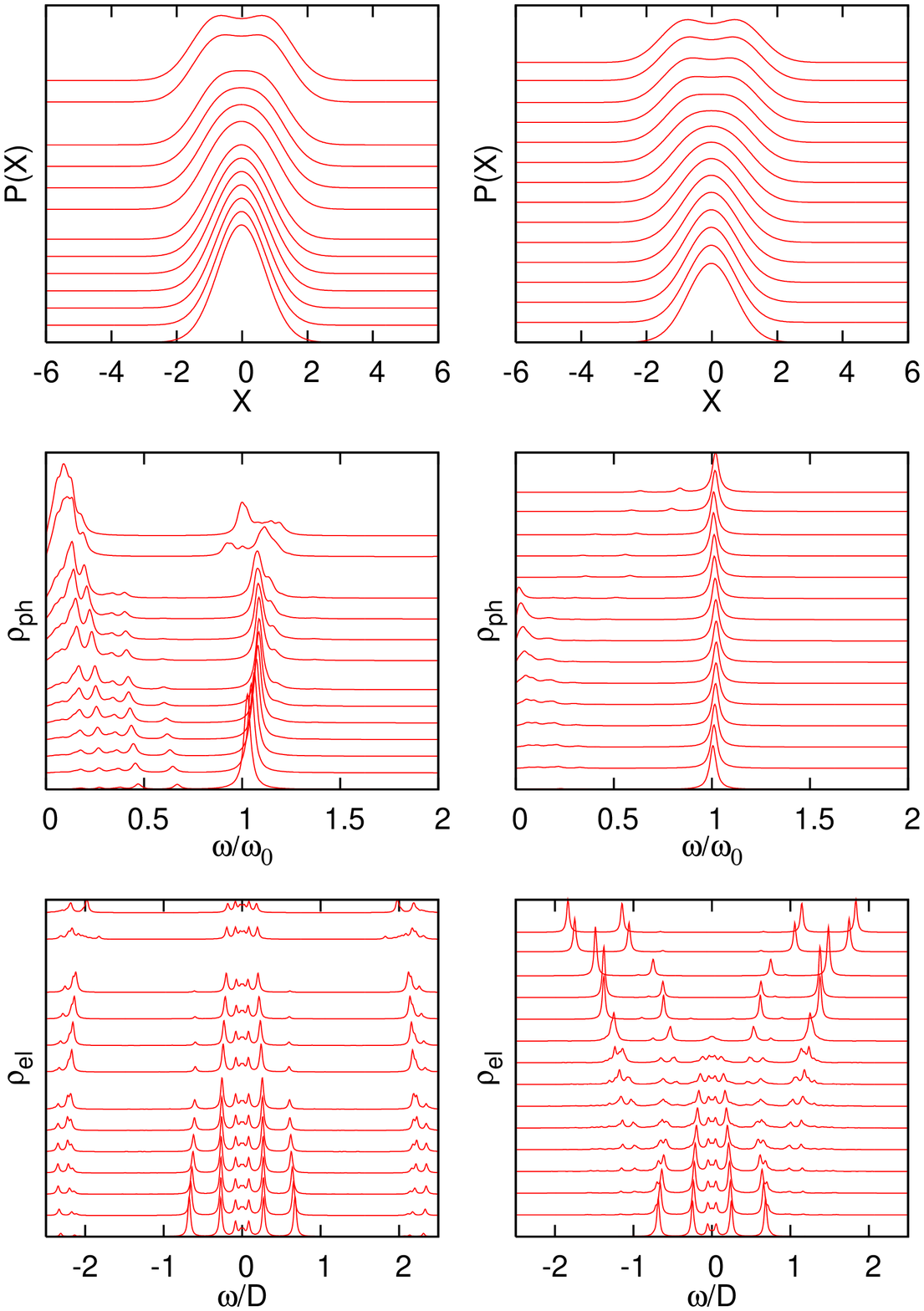}
\end{center}
\caption{(color online) DMFT data in the antiadiabatic regime $\ad=2.0$ spinless
(panels on the left) and
$\ad=4.0$ spinful (panels on the right).
In each panel the various curves refer to different value of
$\lambda$ spanning from  $0.4$ to $6.5$ in the spinless case
and from $0.2$ to $3.0$  in the spinful case
and are shifted according $\lambda$ value.
The first line show the phonon PDF, the central line the phonon DOS
and bottom line the  electronic DOS.}
\label{fig:data_antiad}
\end{figure*}

In this section we briefly anticipate the main results obtained through ED solution
of the full DMFT equations.
We compare spinless and spinful fermions organizing the results according to the
degree of adiabaticity.
The results are summarized in Figs. \ref{fig:data_adiab} and
\ref{fig:data_antiad}, respectively referring to  the adiabatic and
anti-adiabatic regime.
In each figure the left column are dedicated to the spinless fermion case, and the right one to the spinful system.

As we discuss in Sec. \ref{sec:adiabatic}, in strong coupling, we
can introduce relations between the parameters $\lambda$ and $\ad$
which allow for a direct comparison. For the adiabatic parameter we have
$\ad_{spinful} = 2 \ad_{spinless}$. We choose therefore
$\ad=0.1$ ($\ad=0.2$) as representative of the spinless (spinful) 
adiabatic regime  and $\ad=2.0$ ($\ad=4.0$) for the
antiadiabatic regime. 

Let us now qualitatively describe the behavior of the model in these two
regimes by comparing the phonon PDF (First row in  Figs. \ref{fig:data_adiab},
\ref{fig:data_antiad}),  phonon DOS (Central Row in Figs. \ref{fig:data_adiab},
\ref{fig:data_antiad}) and electronic DOS (Bottom row in  Figs.
\ref{fig:data_adiab}, \ref{fig:data_antiad}). Let us remark again that the
discretization of the electronic Hilbert space of the bath allows to extract
only  gross features from the spectral properties of both electron and phonon.

We first discuss phonon properties through an analysis of the evolution of
the phonon  PDF and DOS. 
The PDF has a qualitatively similar behavior in adiabatic and
antiadiabatic cases for spinless and spinful cases: increasing e-ph
coupling the PDF becomes bimodal signaling a polaronic or bipolaronic
phase \cite{Millis-adiab,Max1}.
However the value of e-ph coupling for which the bimodal behavior establishes
is much higher in the antiadiabatic case.

The phonon DOS shows instead a rather strong qualitative dependence on the
adiabatic parameter. Approaching the polaronic region in the adiabatic
regime we observe a softening of the phonon frequency. In the anti-adiabatic
case, instead, we have a transfer of spectral weight from high to low frequency 
while the position of the resonances are roughly unchanged.
In this regard a qualitative difference between the spinless and the spinful
case appears at low energies: in the former case the low energy peak
disappears at strong coupling.

The most important  qualitative difference between the  spinless and the
spinful case is the behavior of the electronic DOS, and it is more evident in the
anti-adiabatic case.
In the spinful case a sharp Metal-Insulator Transition (MIT)
transition, signaled by the opening of a gap in the electron DOS,
is observed for a critical  value of the
coupling which is not dramatically dependent on the adiabatic parameter.
Conversely in the spinless case no sharp MIT takes place, even well inside the
polaronic region.

\begin{figure}[htbp]
\begin{center}
\includegraphics[scale=0.33,angle=270]{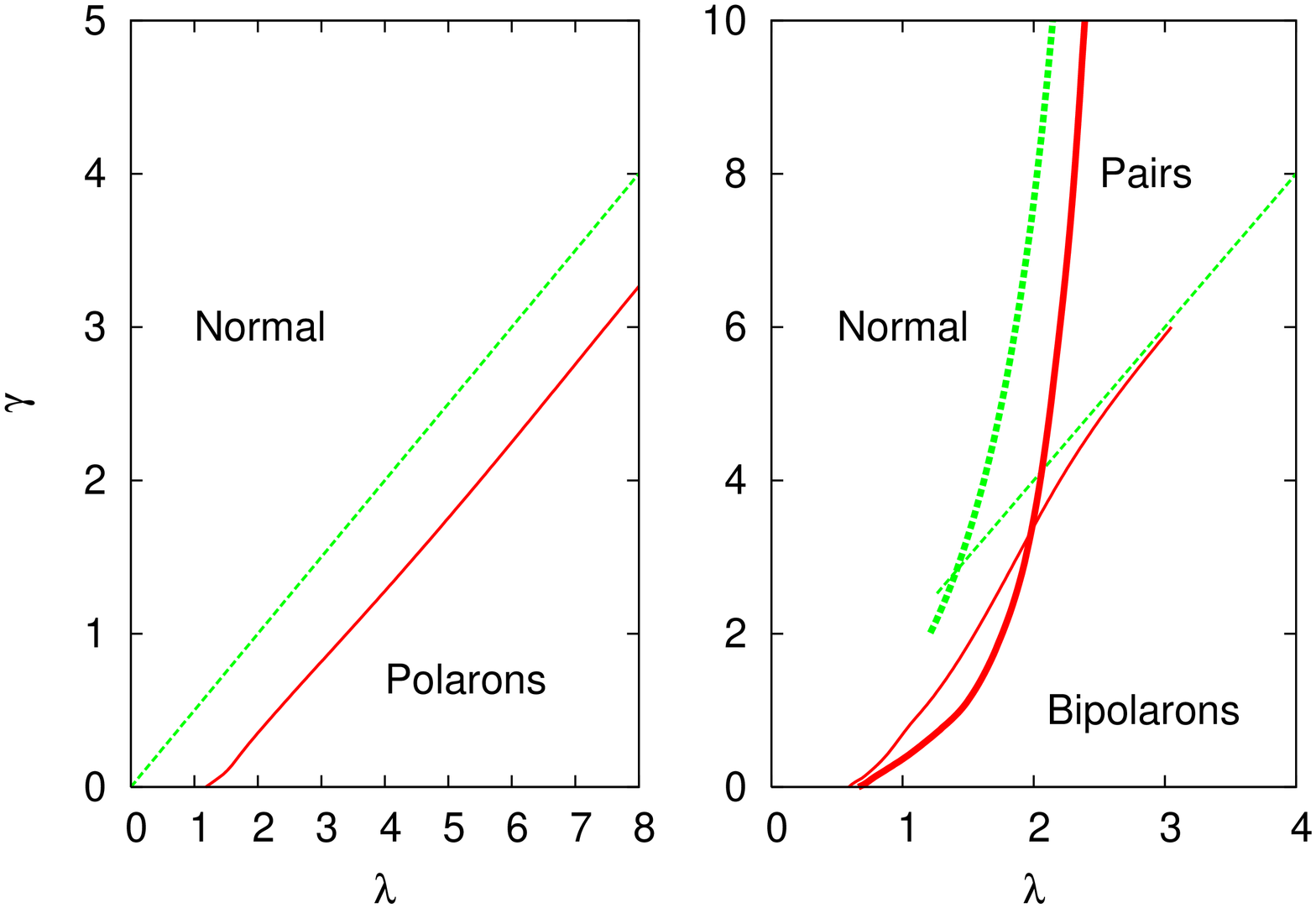}
\end{center}
\caption{(color online) Phase diagrams of the spinless (left) and spinful (right)
Holstein model at half filling. Solid lines: numerical results from DMFT, 
dashed lines:
approximations. Left panel: the bold line is the polaron crossover from bimodality
of $P(X)$ and the dotted line is the anti-adiabatic estimate $\alpha^2>1$ for the polaron
crossover. \\
Right panel:bold curve is the bipolaronic MIT from vanishing of $Z$,
thin solid line the polaron crossover, bold dotted line is the anti-adiabatic
prediction for bipolaronic MIT (Eq. (\ref{eq:MIT_LF})),  light dotted line is
the anti-adiabatic estimate  $\alpha^2>1/4$ for the polaron crossover.}
\label{fig:PD}
\end{figure}

All the above observations  can be summarized in the phase diagrams of Fig.
\ref{fig:PD}.
The points in which the  phonon PDF becomes bimodal are used to draw a line which
represents an estimate of the polaron crossover region. This line is apparently
strongly $\ad$ dependent in both the spinless and spinful case.
Beyond this line a polaronic (bipolaronic) regime is attained in the spinless (spinful) case.
In the spinful case, we can also draw a MIT line, which separates a normal phase
from a paired insulating phase\cite{Max1}.  Notice that in the
antiadiabatic regime, where the Holstein model is approaching a purely electronic
attractive Hubbard model, a pair has not necessarily a well definite associated
polarization therefore at very large value of $\ad$ we can have pairs without
bipolaronic behavior even in the Holstein model.
In the following sections we will analyze separately the adiabatic and the non
adiabatic regimes of the system, discussing the numerical results in comparison
with different analytical approaches suitable for the two opposite regimes.

\section{Adiabatic regime}
\label{sec:adiabatic}
\subsection{Adiabatic limit and BO approximation}

We briefly describe, as a starting point of the Born-Oppenheimer (BO) procedure, the
adiabatic limit in which $\ad \rightarrow 0$ keeping
$\lambda$ fixed. This limit has been thoroughly studied in  Ref.
\cite{Millis-adiab}, and here we introduce a slightly
different, yet equivalent, formulation.

In the adiabatic limit the phonon displacement becomes a classical variable,
therefore the HIM (\ref{eq:Anderson_Holstein}) can be written as
\begin{eqnarray}
\label{eq:HAnderson-adiab}
H &=& -\sum_{k,\sigma} V_k ( c^{\dagger}_{k,\sigma} f_{\sigma} + h.c.)
+ \sum_{k,\sigma} E_k c^{\dagger}_{k,\sigma} c_{k,\sigma} \nonumber\\
&-&
g^\prime n_0 X + \frac{1}{2}k X^2.
\end{eqnarray}
where $g^\prime=g\cl/\sqrt{2}$ and $\cl=\sqrt{\hbar/M\omega_0}$
is the harmonic oscillator characteristic length.

\begin{figure}[htbp]
\begin{center}
\includegraphics[width=6.7cm,height=2cm]{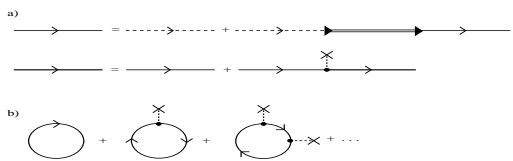}
\end{center}
\caption{a) The equation for $G_0$ (thin line) and $G$ (bold line). Dashed
line is the single site impurity propagator ($1/\omega$) bold arrow is the
hybridization constant $V_\kvec$ double line the bath propagator and x-type
insertion the scattering with static displacements field.
b) The diagrams expansion of the adiabatic potential.}
\label{fig:adiabatic-diagrams}
\end{figure}

This Hamiltonian represents an impurity electron which undergoes multiple
scattering, as depicted in Fig. \ref{fig:adiabatic-diagrams} a), and it can
jump on
the conduction band via the first term of (\ref{eq:HAnderson-adiab}). The
Green's function is immediately written as
\begin{equation}
\label{eq:G-adiab}
G(\omega)= \sum_l w_l \frac{1}{G_0^{-1}(\omega)+g^\prime X_l}
\end{equation}
where $l$ labels all possible value of $X$ giving the same ground state
energy and $w_l$ are the corresponding weights\cite{NoteKF,FJS}.
These values can be obtained by minimizing the total ground state
energy which we call adiabatic potential \cite{NoteKF2,BrandtMielsch,ChungFreericks}
\begin{eqnarray}
V(X)&=&\frac{1}{2}k X^2-\frac{g^\prime}{2} |X|\nonumber\\
&&-\frac{2s}{\beta}\sum_n
\log \left ( \frac{G^{-1}_0(i\omega_n)+g^\prime X}{i\omega_n+g^\prime X}\right )
\label{eq:E-adiab}
\end{eqnarray}
where $s$ is spin degeneracy.

In formula Eq. (\ref{eq:E-adiab}) we have found useful to separate the
contribution in absence of hybridization (
first line of Eq. (\ref{eq:E-adiab})) from a remainder
(last line). The latter term can be obtained easily
trough the linked cluster theorem as depicted graphically in
fig.\ref{fig:adiabatic-diagrams} b)

The equation for the extrema of the adiabatic potential
$X_m$ is
\begin{equation}
\label{eq:X0-adiab}
X_m = \frac{s g^\prime}{k\beta} \sum_n \frac{1}{G^{-1}_0(i\omega_n)+g^\prime X_m} 
\end{equation}
The self-consistency condition (\ref{eq:self-cons}) together with
Eqs. (\ref{eq:G-adiab}) and  (\ref{eq:X0-adiab}), completely solves the problem.
Notice from Eq. (\ref{eq:X0-adiab}) the correspondence  between the spinless and and the spinful case
upon rescaling $\lambda=g^{\prime 2}/k t$ to $\lambda/2$ in the latter case.

In the Bethe lattice case
at half filling it can be shown that for $\lambda<\lambda_c$, where $\lambda_c=3\pi/(8
s)$,  Eq. (\ref{eq:X0-adiab}) has only one solution while for $\lambda > lambda_c$,
three solutions exist:  two stable
($X_l\ne 0$ and $w_l=1/2$) and one unstable ($X=0$) \cite{Millis-adiab}.
Eq. (\ref{eq:G-adiab}) can be recast as an average over
the phonon PDF which is a single $\delta$-function for $\lambda<\lambda_c$ and
splits in two symmetric $\delta$-functions for $\lambda>\lambda_c$.

A MIT occurs at a larger coupling
$\lambda_{MIT} = 1.328/s$ as a vanishing of the DOS at the Fermi level
\cite{Millis-adiab}.

As a final remark we mention the similarity between the adiabatic limit of
the Holstein model and the DMFT solution of the Falicov-Kimball model
\cite{BrandtMielsch} as already
pointed out in Ref. \cite{FJS}. This similarity is evident once we consider the
Coherent Potential Approximation (CPA) form of the Green's function
(\ref{eq:G-adiab}) and compare it with,e.g., that
given in Ref. \cite{ChungFreericks}.

The BO procedure goes on by quantizing the adiabatic potential after
adding the phonon kinetic energy contribution. Introducing the scaled
variable $u=g^\prime X/D\sqrt{s\lambda}$ the BO phononic hamiltonian reads
\beq
\label{eq:BO}
H_{BO}=-\frac{\ad}{2s}\frac{d^2}{d u^2}+s V(u).
\eeq
and $V(u)$ is given by Eq. (\ref{eq:E-adiab}) in the variable $u$.
Notice that the spinful BO hamiltonian
maps onto twice the spinless one upon
rescaling $\lambda/2 \rightarrow \lambda$
and $2\ad \rightarrow \ad$.

Whether the gradient term represents the most relevant contribution from
quantum fluctuation of phonons can be questionable. A more general non-local
contribution to the adiabatic potential arises in the effective phonon action
as it is discussed in Ref. \cite{pata}. As it is discussed there this non local
part may affects the determination of the phonon properties. However we decide
to pursuit the way of simplicity and discuss the BO approximation in view of
comparison with ED data.

While phonon  properties are immediately obtained at this stage from the
solution of the one-dimensional anharmonic system of Hamiltonian  Eq.
(\ref{eq:BO}), electronic properties must account non-trivially
for the tunneling of phonon coordinates.

The simplest way to describe electrons coupled to a tunneling
system is to map it onto a two level system. In our model this can be
accomplished by changing the basis (operators $a$) from that of the
harmonic oscillator to the that defined by the solution of (\ref{eq:BO}).
Then projecting
out all the states but the first two ($\vert +\rangle,\vert -\rangle$) we get the following two
state projected model (TSPM):
\begin{eqnarray}
\label{eq:TSPM-definition}
H&=&-\frac{2}{s}\sum_\sigma\e\left(f_\sigma^+f_\sigma-\frac{1}{2}\right)\sz -\D\sx
+\sum_{k,\sigma}E_k c^\dagger_{k,\sigma}c_{k,\sigma}+
\nonumber\\
&+&\sum_{k,\sigma} V_k \left( f_\sigma^\dagger c_{k,\sigma} + c^\dagger_{k,\sigma} f_\sigma\right),
\end{eqnarray}
where $\sigma_x$ and $\sigma_z$ are two Pauli matrices in the space spanned by $\vert + \rangle$
and $\vert - \rangle$ and the quantities $\epsilon$ and $\Delta$ are given by
\begin{eqnarray}
\label{eq:TSPM-parameters}
 \epsilon &=& g\frac{s}{2} \langle+\vert a+a^\dagger\vert -\rangle\\
 \Delta &=& \frac{\omega_0}{2}(\langle +\vert a^\dagger a\vert +\langle - \rangle -\vert a^\dagger a\vert -\rangle)
\end{eqnarray}
The latter quantity $\Delta$ as a clear meaning as a tunneling frequency
between the two phononic states.
A similar model was introduced in Ref. \cite{pata} to study the strong
coupling limit of the Holstein model.
Here we remark that we use this model as a tool to gain physical insight in
the analysis of numerically exact output of DMFT.

\begin{figure}[htbl]
\begin{center}
\includegraphics[scale=0.33,angle=270]{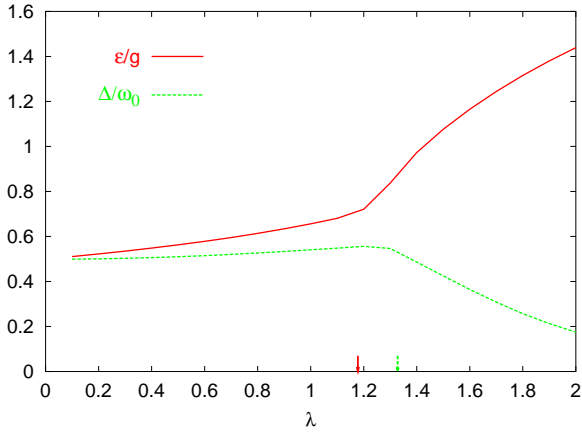}
\end{center}
\caption{(color online) Parameters of the TSPM in the spinless case for
$\ad=0.1$. The spinful case is simply obtained by taking  
$\lambda_{spinful}=\lambda_{spinless}/2$. The adiabatic polaronic transition
$\lambda_c$ is marked by a solid arrow while the adiabatic MIT
is marked by a
dashed arrow.}
\label{fig:epsdelta}
\end{figure}

A sample plot of the parameters of this model as obtained from the adiabatic
limit of the Hamiltonian (\ref{eq:themodel}) is reported in fig. \ref{fig:epsdelta}.
At weak coupling we obtain $\e \simeq g\frac{s}{2}$ and  $\Delta \simeq \frac{\omega_0}{2}$.
At strong coupling instead
 $\e$ scales as $\lambda /4$ while $\Delta$ vanishes exponentially, even if
it never becomes strictly zero for any finite value of e-ph coupling in both spinless and spinful case.

There are two limits in which the  TSPM reproduces exactly the original Holstein
model within DMFT: the weak coupling  and the adiabatic limit.  In
the former case the projection of the phonon space has no relevance therefore the
TSPM reproduces the perturbation expansion developed (in the limit of infinite bandwidth)
in the classical Ref. \onlinecite{Engelsberg}.
The adiabatic limit is instead recovered as $\Delta\rightarrow
0$. No phonon tunneling occurs and the model can be solved exactly by CPA
recovering the solution of Ref. \onlinecite{Millis-adiab}.

\begin{figure}[htbp]
\begin{center}
\includegraphics[scale=0.33,angle=270]{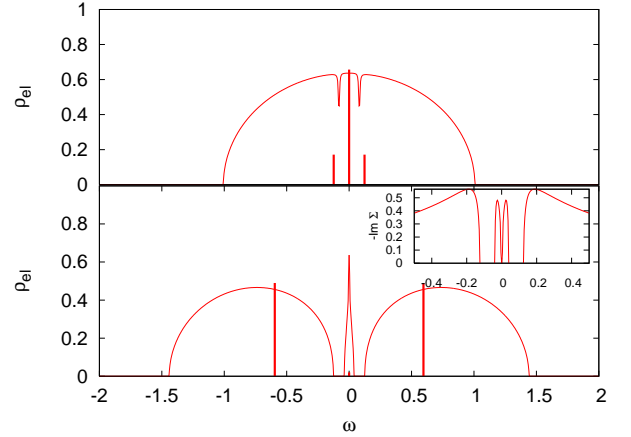}
\end{center}
\caption{(color online) The spectral function of the spinless TSPM in the zero hybridization case
(bold lines)  and in the CPA approximation (thin line). The upper panel
refers to a typical weak-coupling situation ($\lambda=0.1$,$\ad=0.1$) while 
the lower panel presents a  strong-coupling case ($\lambda=1.6$,$\ad=0.1$). 
In the inset of the lower panel $-Im \Sigma$ is shown.}
\label{fig:CPAspinless}
\end{figure}

\begin{figure}[htbp]
\begin{center}
\includegraphics[scale=0.33,angle=270]{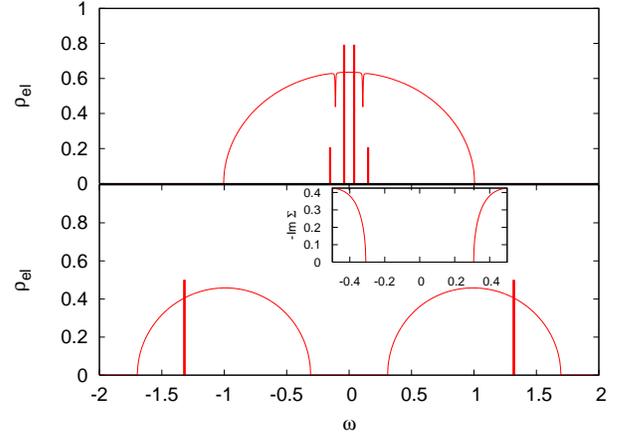}
\end{center}
\caption{(color online) The spectral function of the spinful TSPM in the zero
hybridization case (bold line) and in the CPA approximation (thin
line). The upper panel shows a weak-coupling result ($\lambda=0.2$,$\ad=0.1$),
the lower panel a strong coupling one ($\lambda=1.9$,$\ad=0.1$).
In the inset of the lower panel $-Im \Sigma$ is shown.}
\label{fig:CPAspinful}
\end{figure}

To study the effect of phonon tunneling events in strong coupling
it is useful to consider the TSPM in absence of hybridization
($V_{\kvec}=0$). Taking into account the self-consistency condition
 (\ref{eq:self-cons}), this is the atomic ($D=0$) case for the lattice Hamiltonian
(\ref{eq:themodel}).
Let us first consider the
simplest spinless case where the eigenstates are given by (see also
Appendix \ref{app:App_pert} for details)
\begin{eqnarray}
\label{eq:eigvecspinless}
|v^+_\s\rangle =|0\rangle \otimes
 \frac{1}{\sqrt{2\l(\l-\s\e)}}\begin{pmatrix}
 \D \\ \e-\l\s
 \end{pmatrix} \qquad
 \nonumber \\
|v^-_\s\rangle =f^+|0\rangle \otimes
 \frac{1}{\sqrt{2\l(\l+\s\e)}}\begin{pmatrix}
 \e+\l\s \\ \D
 \end{pmatrix},
\end{eqnarray}
where $\s=\pm 1$ labels the phonon "spin" and $|0\rangle$ is the Fock vacuum for
the impurity. The two-component spinors live in the space defined by
$\vert +\rangle$ and $\vert -\rangle$.
The eigenvalues are in this case  doubly degenerate:
\begin{equation}
\label{eq:eigvalspinless}
E_\s^\sprime  = \s \sprime \l
\end{equation}
where $\sprime,\s=\pm 1.$  and $\l^2 = \e^2+\Delta^2$.

The eigenstates $|v^-\rangle$ and $|v^+\rangle$ describe respectively the presence and
the absence of an electron on the impurity, and they are associated to different phononic
states. The phononic tunneling $\Delta$ allows for the existence of quantum {\it
defects} on the impurity i.e. states in which the impurity is occupied (empty)
and the deformation is the one adiabatically associated to the absence (presence)
of the electron,e.g., $|v^{+}_{+}\rangle$ and  $|v^{-}_{-}\rangle$ . The adiabatic case
is recovered once $\Delta=0$ so that  only $|v^{+}_{-}\rangle$ and
$|v^{-}_{+}\rangle$ survive.

The atomic ($V_k=0$) Green's function can be easily found to be
\beq
\label{eq:GatBOCPAspinless}
G_a(\o) = \frac{1}{2}\sum_{\s=\pm}
\left( \frac{\e^2}{\l^2}\frac{1}{\o+2\l\s} +
\frac{\D^2}{\l^2}\frac{1}{\o}\right)
\eeq
$G_a(\omega)$ has a pole at $\omega=0$ induced by phonon tunneling, whose weight in fact
vanishes  as $\Delta\rightarrow 0$, accompanied by
two resonances at $\pm 2\l$, as depicted in Fig. \ref{fig:CPAspinless}.
The zero energy peak is due to transitions in which both charge and phonon "spin"
change while the side peaks take into account charge transfer in a frozen phonon "spin".
In this sense the side peaks are adiabatic features which survive when phonon tunneling
$\Delta\rightarrow 0$.

In the spinful case denoting with $|v\rangle$ the singly occupied states and with
$|u\rangle$ the doubly  occupied or empty states we obtain
(see Appendix\ref{app:App_pert})
\begin{eqnarray}
\label{eq:eigvecspinful1}
|u^+_\s\rangle =|0\rangle \otimes
 \frac{1}{\sqrt{2\l(\l-\s\e)}}\begin{pmatrix}
 \D \\ \e-\l\s
 \end{pmatrix} \qquad
 \nonumber \\
|u^-_\s\rangle =f^+_\uparrow f^+_\downarrow|0\rangle \otimes
 \frac{1}{\sqrt{2\l(\l+\s\e)}}\begin{pmatrix}
 \e+\l\s \\ \D
 \end{pmatrix},
\end{eqnarray}
\begin{eqnarray}
\label{eq:eigvecspinful2}
|v^\beta_+\rangle =f^+_\beta|0\rangle \otimes\frac{1}{\sqrt{2}}
\begin{pmatrix}
 1 \\ -1
 \end{pmatrix} \qquad
 \nonumber \\
|v^\beta_-\rangle =f^+_\beta|0\rangle \otimes\frac{1}{\sqrt{2}}
\begin{pmatrix}
 1 \\ 1
 \end{pmatrix}.
\end{eqnarray}
The eigenvalues are
\begin{eqnarray}
\label{eq:eigvalspinful}
E_\s^\sprime (u) = \s\sprime \l
\nonumber \\
E_\s^\sprime (v) = \s\sprime \D
\end{eqnarray}
each level is doubly degenerate and $\sprime,\s=\pm 1$.

Notice that the empty or doubly occupied impurity states $\vert u \rangle$
 are associated to a finite deformation, while the singly occupied states
are associated to a tunneling phonon state (eigenstates of $\sigma_x$).

In this case a single particle excitation has always non zero energy as it can be
seen by the Green's function
\beq
\label{eq:GBOCPAspinful}
G_a(\o)=\frac{1}{2}\sum_{\s=\pm}\frac{1}{2\l}\left(
\frac{\l-\D}{\o+\s(\l+\D)}+\frac{\l+\D}{\o+\s(\l-\D)}\right)
\eeq
depicted in Fig. \ref{fig:CPAspinful}. The
most striking with respect to the spinless case (\ref{eq:GatBOCPAspinless}) is the
absence of the zero frequency pole. In the spinful case the tunneling of the
phonon $\Delta$ is always associated to a finite energy transition, and it only splits the
finite frequency poles associated to the transition from singly  to
the empty or doubly occupied ones.

To analytically span from the strong ($V_\kvec \rightarrow 0$) to the  weak
($g\rightarrow 0$) coupling regimes of the equivalent impurity Hamiltonian it
is useful to devise a Born-Oppenheimer Coherent Potential Approximation (BOCPA) scheme.
Starting from the Green's function for $V_k=0$
(\ref{eq:GatBOCPAspinless}) for the spinless case and (\ref{eq:GBOCPAspinful}) for the spinful
one, we notice that in both cases $G_a(\o)$ can be written ad sum of two contribution
$G_a(\o)=(1/2)(G_{a,+}(\o)+G_{a,-}(\o))$ where
($\pm$) label a phonon state as in
(\ref{eq:eigvecspinless}) and (\ref{eq:eigvecspinful1}).
Then we write the propagator in presence of hybridization as
\beq
\label{eq:GBOCPA}
G(\o)=\frac{1}{2}\sum_{\s=\pm}\frac{1}{G^{-1}_{a,\s}(\o)-\Sigma_{ibr}(\o)}
\eeq
where $\Sigma_{ibr}=\sum_k V_k^2/(\omega-E_k)$
and takes into account the hopping of the electron from the impurity to the
conduction states (see Fig. \ref{fig:adiabatic-diagrams}).
BOCPA assumes that the tunneling states of the phonon in the adiabatic potential
remain unaltered during an hybridization event.
The results depicted in figs.
\ref{fig:CPAspinless} and \ref{fig:CPAspinful}. As in standard CPA at each local
level is associated a band but in contrast with the CPA for the Hubbard model,
our  BOCPA gives a Fermi liquid solution in the spinless case for every value of the
coupling.
The low-energy band arising from the zero energy pole in the zero
hybridization limit is indeed {\it coherent}. This can be easily realized by
analysis of  the self-energy. When $\omega\rightarrow 0$ the propagator
the spinless propagator defined by  (\ref{eq:GatBOCPAspinless}) and (\ref{eq:GBOCPA}) is
dominated by the zero-energy pole of $G_a$  (\ref{eq:GatBOCPAspinless}) and
consequently the self-energy obtained through (\ref{eq:sigma}) is purely real.

Conversely in the spinful case a MIT
transition similar to that of the Hubbard model in the CPA approximation is observed at a critical value
of $\lambda$ \cite{noteMITCPA}. In both spinless and spinful case the weak
coupling limit of BOCPA (thin lines of upper panels of Figs.
\ref{fig:CPAspinless} and \ref{fig:CPAspinful}) bears strong
resemblance with the classical result of Ref. \onlinecite{Engelsberg}.
Finally we emphasize that we recover the adiabatic solution of Ref. \onlinecite{Millis-adiab}
as $\ad \rightarrow 0$ for finite $\lambda$ as $\Delta\rightarrow 0$. In
this case the BOCPA is exact and gives the Green's function of Eqs. (\ref{eq:G-adiab}).
However the BOCPA procedure is certainly affected by serious problems
approaching the MIT in the spinful case. A more careful treatment of
the low-energy part of the Green's function has been performed in this case
in Ref. \cite{pata}, where it has been observed a MIT scenario similar to the
half-filled the Hubbard model, i.e., a quasi-particle peak that shrinks to zero
width approaching a critical value of $\lambda$,
 as it will be discussed at the end of the next section.
In the spinless case instead
a resonance is present at zero energy in the
even within a CPA approach. It is not associated to a Kondo effect
but rather to phonon tunneling which drives charge fluctuations. On the
other hand a Kondo like behavior can be ascribed to the bipolaron or pair
formation (spinful case) and should be treated either numerically as in
our case or with a more appropriate theory as in Ref. \onlinecite{pata}.

Now let us briefly discuss which kind of processes are left out from a
BOCPA theory. For the sake of clarity here we only present the main
results of a perturbative treatment up to second order in the hybridization $V_k$,
and we refer to Appendix \ref{app:App_pert} for more details.

\begin{figure}[htbp]
\begin{center}
\includegraphics[width=6.7cm,height=2cm]{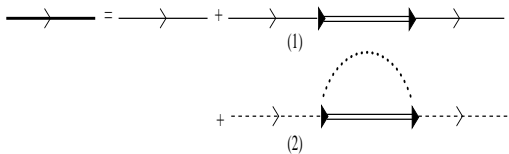}
\end{center}
\caption{The second order perturbation expansion in $V_k$ for the impurity propagator.
Symbols are defined in the Appendix \ref{app:App_pert}}
\label{fig:pertVk}
\end{figure}
The electron Green's function can be written at second order in hybridization $V_k$ as
$G(\omega)=G_1(\omega)+G_2(\omega)$ where $G_1$ is a term which is resummed in BOCPA scheme and
$G_2$ is an extra term coming from processes in which the phonon ``spin'' changes during the hybridization
process. The two terms are represented diagrammatically in Fig. \ref{fig:pertVk}.
$G_2$ reads:
\beq
\label{eq:eqG2}
G_2(\omega)=\sum_\s \tilde{g}_\s (\omega)\Sigma_\s ^\prime(\omega)\tilde{g}_\s (\omega)
\eeq
where
\beq
\Sigma_\s ^\prime(\omega)=\sum_k \Theta(-\s E_k)\frac{|V_k|^2}{\omega-E_k+2\l \s}
\eeq
and the step function $\Theta(-E_k)$ represents the Fermi function at zero temperature.
In Eq. (\ref{eq:eqG2}) $\tilde{g}_\s (\omega)$ is in the spinless case
\beq
\tilde{g}_\s (\omega)=\frac{\e \Delta}{\l ^2}
\left ( \frac{1}{\omega+2\l \s}-\frac{1}{\omega}  \right )
\eeq
while in the spinful case is given by
\beq
\tilde{g}_\s (\omega)=\frac{\e }{\l }
\left ( \frac{1}{\omega+\s (\l + \Delta )}-\frac{1}{\omega+\s (\l - \Delta )}  \right ).
\eeq
As discussed in \ref{app:App_pert}, the $\tilde{g}$ propagators do not carry charge,
as they are correlation functions of the phonon "spin", and we will refer to them
as  $f-$spinons. By inspecting the structure
of the Dyson's equation (\ref{eq:eqG2}) we see that the insulating
character is preserved after the inclusion of $G_2$ terms in the spinful
case. In the spinless case a modification of the zero energy pole arises
from Eq. (\ref{eq:eqG2}) but it is still present also after inclusion of the
$f-$spinon terms. We notice however that approaching zero energy
the $f-$spinon propagator becomes $\s $ independent in the spinless case.
Assuming particle-hole symmetry $\sum_\s \Sigma_\s ^\prime(\omega=0)=0$
which leads to a vanishing contribution of the $f-$spinon diagrams at zero
energy leading to the validity of the Luttinger's theorem even within BOCPA for
the spinless case.

\subsection{Discussion of the results in the adiabatic regime}

\begin{figure}[htbp]
\begin{center}
\includegraphics[scale=0.33,angle=270]{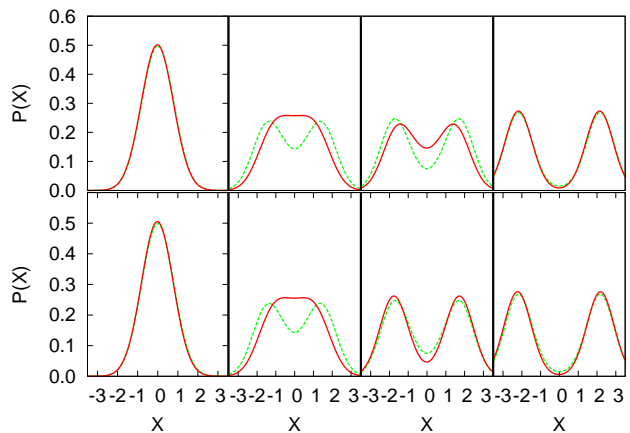}
\end{center}
\caption{(color online) The phonon PDF in the spinless (upper panel) and spinful (lower panel) cases.
Solid curves are DMFT results and dashed lines the BO approximation.
The values of $\lambda$ are in the spinless (spinful case) from left to right:
$\lambda=0.5$($0.25$),$\lambda=1.5$($0.75$),$\lambda=1.7$($0.85$),
$\lambda=2.2$($1.1$)}
\label{fig:PXvsBO}
\end{figure}

Fig. \ref{fig:PXvsBO} presents a comparison between the BO approximation and the
DMFT results for the phonon PDF.
The anharmonicity due to e-ph interaction increases as the
coupling increases leading first to a non-gaussian and finally to a bimodal PDF.
This behavior signals the appearance of static distortions, even if we are neglecting
any ordering between them.
We can estimate a ``central line'' of the region in which the polaron crossover
takes place according to the values of $\lambda$ and $\ad$ at which phonon PDF becomes bimodal.
From Fig. \ref{fig:PXvsBO} is evident that BO approximation works well in
{\it both} the metallic and the polaronic regimes. The reason for the accuracy of the
BO procedure in the polaronic regime is that, contrary to its usual
implementation in the weak e-ph coupling\cite{russianBO},
here we  take into account the anaharmonicity
through Eq. (\ref{eq:BO}) in a non perturbative way.
However, BO does not accurately reproduce the  phonon PDF around the polaron crossover.
In this case electron and phonon states are  strongly entangled, and cannot be approximated
properly by a disentangled BO state.
By a comparison of the spinless and spinful cases in fig. \ref{fig:PXvsBO} we
see that the occurrence of the MIT does not influence much the differences between
full DMFT and BO, which are in both cases relevant near the polaron crossover.

\begin{figure}[htbp]
\begin{center}
\includegraphics[scale=0.33,angle=270]{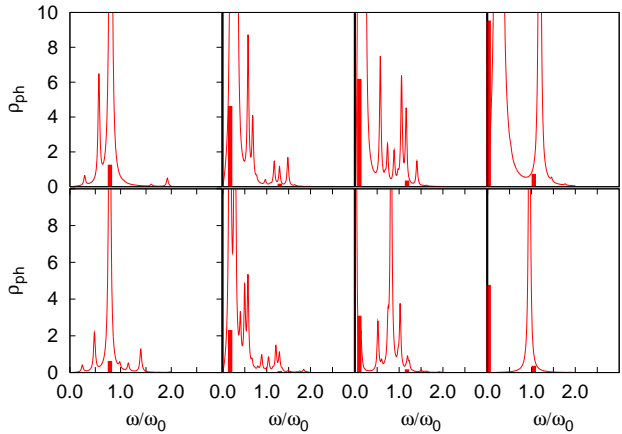}
\end{center}
\caption{(color online) Phonon DOS in the spinless (upper panel) and spinful (lower panel) cases.
Solid curves are DMFT results, bold impulses the BO approximation.
The values of $\lambda$ are the same as in Fig. \ref{fig:PXvsBO}
}
\label{fig:DOSPHvsBO}
\end{figure}

\begin{figure}[htbp]
\begin{center}
\includegraphics[scale=0.33,angle=270]{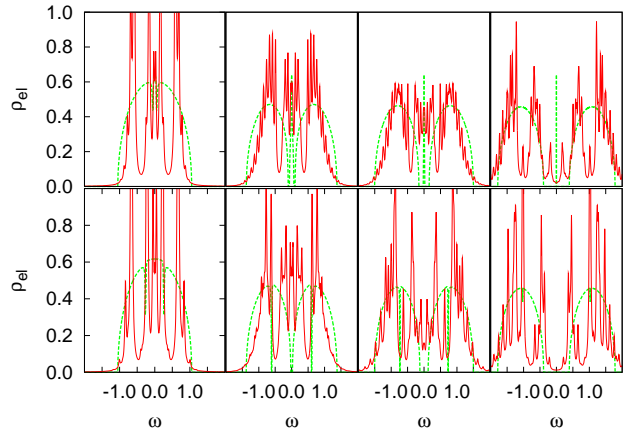}
\end{center}
\caption{(color online) DOS in the spinless (upper panel) and spinful (lower panel) cases.
Solid curves are DMFT results, dashed lines the BOCPA approximation.
The values of $\lambda$ are the same as in Fig. \ref{fig:PXvsBO}}
\label{fig:DOSvsBO}
\end{figure}

This is not the case of the phonon DOS shown in Fig. \ref{fig:DOSvsBO}. In
this case a qualitative difference between the spinless and the spinful case
appears at low energy as already noticed discussing Figs. \ref{fig:data_adiab} and
\ref{fig:data_antiad}: In the spinful case and above the MIT
(rightmost picture in the lower row of Fig.\ref{fig:DOSPHvsBO}),
the low-energy peak of the phonon DOS completely disappears.
This behavior cannot be predicted by BO approximation which gives undamped
phonon peaks approaching zero with {\it increasing} spectral weight.
An explanation of this behavior needs an anti-adiabatic approach which will be
described in the next section.

The data for the phonon DOS appearing in Figs. \ref{fig:data_adiab} and
\ref{fig:DOSPHvsBO} can be qualitatively compared to those obtained by Numerical
Renormalization Group (NRG) solution of the impurity model (See, Fig. 5 of Ref. \onlinecite{Bulla}).
The main differences between our ED results and NRG seems to be the pole at
$\omega_0$ which is present in NRG for all the coupling strengths shown,  while in
our analysis this pole softens and finally disappear at the MIT.
We ascribe this different behavior to the effect of the high energy part of
electronic spectrum, which is treated approximately by NRG but it may strongly
influence  the phonon properties. This is easily realized by analyzing the
simple bubble diagram for the phonon self-energy, in which both
low- and high-energy  electronic scales contribute in the same way the low-energy
part of the phonon self-energy.

In Fig. \ref{fig:DOSvsBO} we compare the electronic
DOS from full DMFT with  BOCPA. The different behavior of the spinless and
spinful case is not so evident in this adiabatic regime.
However, comparing the spinless spectrum for $\lambda=1.5$ with the corresponding
spinful for $\lambda = 0.75$, a quasiparticle peak is present in the former case,
while a depletion of low energy part is much more evident in the latter.
At strong coupling the discretization of the bath degrees of freedom
inherent to the ED solution of DMFT
does  not allow us to identify  a well defined quasiparticle peak in the spinless case.
However a more careful analysis of the quasi particle spectral weight \cite{Max1}
shows that no MIT occurs in the spinless case.
Notice that the BOCPA seems to be much closer to DMFT-ED in the spinless than in
the spinful case where the CPA approximation for the electronic degree of
freedom is apparently much less adequate.

\section{Anti-adiabatic regime}
\label{sec:antiadiabatic}
\subsection{Canonical transformation analysis}

While in adiabatic limit the phonon displacement becomes a classical
variable, and we are left with an electronic model which depends
parametrically on it, in the opposite limit ($\ad >> 1$) the roles are
exchanged, and we have a parametrically
fixed electronic charge on a given site. In this regime the most
reasonable starting point is  the Lang-Firsov (LF) canonical transformation\cite{Lang-Firsov}
$U=\exp (S)$, which here we only apply to
the equivalent impurity Hamiltonian (\ref{eq:Anderson_Holstein}).
The generator of the transformation reads
\beq
\label{eq:LangFirsov}
S = -\alpha \sum_\sigma (f^\dagger_\sigma
f_\sigma-\frac{1}{2})(a^\dagger-a),
\eeq
introducing the parameter $\alpha=g/\omega_0$ which is the relevant e-ph coupling
parameter in the anti-adiabatic regime.\cite{storia,depolarone}
Phonon and impurity electron operators are transformed in the
following way:
\beqa
\label{eq:LangFirsovTrasf}
\tilde{a} = \exp(S)a\exp(-S)&=& a + \sum_\sigma (f^\dagger_\sigma f_\sigma-\frac{1}{2})\\
\label{eq:LangFirsovTrasf2}
\tilde f_{\sigma} = \exp(S)f_\sigma\exp(-S)&=&f_\sigma \exp(\alpha (a^\dagger-a))
\eeqa
This canonical transformation diagonalizes the impurity Hamiltonian in the absence
of hybridization by elmininating the e-ph interaction part.
In the spinful case the phonon energy term of (\ref{eq:Anderson_Holstein})
gives rise to the well known bipolaronic  instantaneous attraction.
The hybridization
term of (\ref{eq:Anderson_Holstein}) is modified according to
(\ref{eq:LangFirsovTrasf2}) by acquiring an exponential term in the phonon
coordinates leading to
\beqa
\label{eq:AndersonLF}
e^S H e^{-S} &=& -\sum_{k,\sigma} e^{\alpha (a^\dagger-a)}
V_k (c^{\dagger}_{k,\sigma} f_{\sigma} +
h.c.) +
\nonumber\\
&+&\sum_{k,\sigma} E_k c^{\dagger}_{k,\sigma} c_{k,\sigma}
-2\frac{g^2}{\omega_0} (s-1) n_\uparrow n_\downarrow -
\nonumber\\
&-&\frac{g^2}{\omega_0} \sum_\sigma (\frac{1}{2}- n_\sigma)
 + \omega_0 a^{\dagger}  a,
\eeqa
where $n_{\sigma}=f^{\dagger}_{\sigma} f_{\sigma}$.

Notice that in the anti-adiabatic limit  $\ad \rightarrow \infty$,
 if $\lambda$ is kept
constant $\alpha$ vanishes. In this case spinless electrons
results unrenormalized while spinful electrons are described by an attractive
Hubbard model with $|U|/D=\lambda$.

If we want to proceed with analytical methods, the hybridization term must
be  treated in an
approximate way. Assuming that in the anti-adiabatic limit the impurity density
is constant during the fast motion of the phonon we average out the phonon term
on the displaced phonon ground state. This is the so-called Holstein Lang-Firsov
Approximation (HLFA), which has not to be confused with the exact canonical transformation
treatment (\ref{eq:LangFirsov}).
HLFA gives rise to the exponential
renormalization of the hybridization constants where each $V_k$ of Eq. (\ref{eq:AndersonLF})
is replaced by $V_k\exp(-\alpha^2/2)$. Through the
self-consistency condition (\ref{eq:self-cons}), such a replacement implies
the well known exponential renormalization of the bandwidth $D\exp(-\alpha^2)$.

The main results of HLFA can be summarized as follows:

{\it i)} In the spinless case the HLFA on the Holstein impurity Hamiltonian
generates a non interacting impurity, which is connected with the bath through an
exponentially reduced hybridization;

{\it ii)} In the spinful case the impurity site presents an attractive instantaneous
interaction term of the Hubbard type $U=-2g^2/\omega_0$.

To get the {\it electron} Green's function $G(\omega)$, the explicit action of LF transformation
(\ref{eq:LangFirsovTrasf}) and (\ref{eq:LangFirsovTrasf2})
have to be taken into account. Following Refs. \onlinecite{Ranninger_spectral} and
\onlinecite{Alexandrov-Ranninger}, we obtain in both spinless and spinful cases
\beq
\label{eq:GreenLF}
G(\omega)=e^{-\alpha^2}G_p(\omega)
+\frac{1}{2}\sum_{n\ne0}e^{-\alpha^2} \frac{\alpha^{2|n|}}{|n|!} G_p(\omega-n\omega_0).
\eeq
where $G_p(\omega)$  is the Green's function of an impurity (with a  negative $U$
interaction in the spinful case) hybridized with an exponentially reduced
hybridization to a bath of conduction electrons. The self-consistency condition can
be written explicitly in the spinless case due to the lack of interaction terms on
the impurity:
\beq
\label{eq:Gp_spinless}
G_p(\omega) = \frac{1}{\omega-e^{-\alpha^2}\frac{t^2}{4}G(\omega)}.
\eeq
where $G(\o)$ is the local Green's function of the lattice.
In the spinful case a Lang-Firsov Coherent Potential Approximation (LFCPA)
can be devised for resulting HLFA
negative-U Hubbard model giving
\begin{eqnarray}
\label{eq:Gp_spinful}
G_p(\omega) &=& \frac{1}{2} \left (
\frac{1}{\omega-e^{-\alpha^2}\frac{t^2}{4}G(\omega)-U/2}+ \right.
\nonumber \\
&+& \left. \frac{1}{\omega-e^{-\alpha^2}\frac{t^2}{4}G(\omega)+U/2}
\right ) .
\end{eqnarray}

\begin{figure}[htbp]
\begin{center}
\includegraphics[scale=0.33,angle=270]{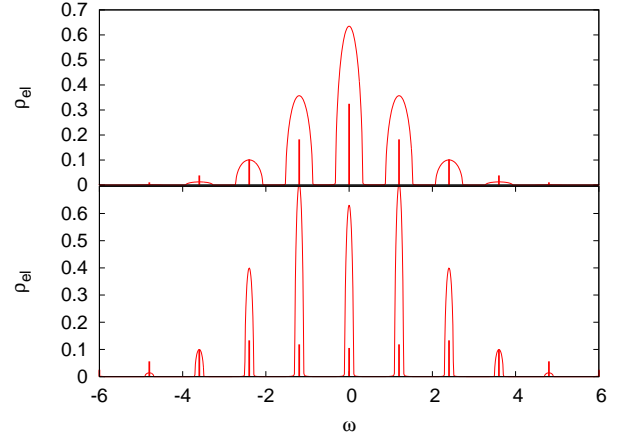}
\end{center}
\caption{(color online) The spectral function in the spinless case in the zero hybridization case
(bold line)  and in the HLFA approximation (thin line). The upper panel
refers to a typical weak-coupling situation ($\lambda=2.7,\ad=1.2$)
while the lower panel shows results for intermediate
coupling ($\lambda=5.4,\ad=1.2$).}

\label{fig:LFspinless}
\end{figure}

\begin{figure}[htbp]
\begin{center}
\includegraphics[scale=0.33,angle=270]{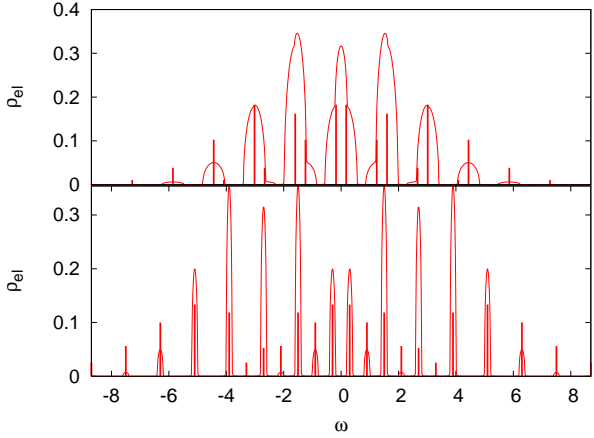}
\end{center}
\caption{(color online) The spectral function of the spinful case in the zero
hybridization case (bold line)  and in the LFCPA approximation (thin
line). The upper panel is dedicated to weak coupling ($\lambda=2.7,\ad=1.2$)
and the  lower panel to 
intermediate coupling
($\lambda=5.4,\ad=1.2$).}
\label{fig:LFspinful}
\end{figure}

A comparison between the zero hybridization (atomic $D=0$) DOS and
results from HLFA and LFCPA results is depicted in figs.
\ref{fig:LFspinless}, \ref{fig:LFspinful}.

Notice that the theory developed here for the Holstein impurity model differs
significantly from that developed directly in the lattice
model\cite{Ranninger_spectral}.
In that case an equation identical to (\ref{eq:GreenLF}) is
recovered for a band of free electrons therefore giving a low-energy {\it
coherent} polaronic band in the spinless case.
It is however easy to show that this form of the spectral function is not compatible with
a $k$-independent self-energy.
Self consistency condition eq. (\ref{eq:Gp_spinless}) which enforces a local self-energy
give rise to a non zero
damping at the Fermi level even in the spinless case.
However as far as $\ad $ becomes larger
$\alpha$ gets smaller reproducing the anti-adiabatic coherent behavior at low
energy in the spinless and the negative-$U$ behavior in the spinful cases.

In the anti-adiabatic regime the HLFA approach gives an estimate of the MIT
by the simple scaling
\beq
\lambda_{MIT}=|U/D|_{MIT}\exp(-\alpha^2)
\eeq
where $|U/D|_{MIT}=2.94$ is the MIT value of the negative-U Hubbard model\cite{Uc1Uc2neg}
which is equal to that of the repulsive model \cite{Uc1Uc2}.
As a function of the adiabatic ratio we get:
\beq
\label{eq:MIT_LF}
\gamma_{MIT}= - |U/D|_{MIT} \frac{\lambda}{2 \log(\lambda/|U/D|_{MIT})}
\eeq
which indicates that the MIT takes place at lower values of $\lambda$
as the adiabatic regime is approached (see Fig. \ref{fig:PD}).
Of course the previous formula has not to be applied in the adiabatic regime.

The phonon PDF can be easily derived within LF approach.
Being the local electron densities parametric variables in the anti-adiabatic
limit the phonon PDF can be written as
\beq
P(X)=\sum_l w_l P_0(X-X_l)
\eeq
where $w_l$ is the probability of having an occupancy $n_l$,
$X_l$ the relative displacements and $P_0(X)$ the ground state PDF of an
harmonic oscillator. $P_0(X-X_l)$ is then the conditional probability of having a
displacement $X$ given a definite occupation $n_l$.
In the spinless case $n_l=0,1$ with equal probability
giving
\beq
\label{eq:PXLFspinless}
P(X)=\frac{1}{2} \left ( P_0(X-X_0)+ P_0(X+X_0) \right)
\eeq
where $X_0=\cl\alpha/\sqrt{2}$.
A definite polarization can be associated to the ground state if the PDF
becomes bimodal. By requiring $d P(X)/d X\vert_{X=0} > 0$, which simply means that
the $X=0$ turns into a local minimum from the maximum it is in weak coupling,
we get the usual anti-adiabatic condition for the existence of a polaronic state, i.e.,
$\alpha^2>1$ (see fig. \ref{fig:PD}).

In the spinful case $n_l=0,1,2$
\beq
\label{eq:PXLFspinful}
P(X)= n_d (P_0(X-2X_0)+P_0(X+2X_0))+(1-2n_d)P_0(X)
\eeq
where $n_d= \langle n_\uparrow n_\downarrow\rangle$ is the site double occupancy.
It is worth noting that in  the insulating state $n_d\simeq1/2$,
and the zero-displacement PDF associated to singly occupied sites is depleted.
This is an example of the
strong coupling dependence of the phonon properties on electronic state.
In contrast with the spinless case the existence of a definite polarization is
now associated to a bipolaronic state.
The condition under which  (\ref{eq:PXLFspinful}) becomes bimodal, is 
\beq
exp(-2\alpha^2)(4\alpha^2-1) \ge \frac{1-2n_d}{2n_d}.
\eeq

An estimate for the bipolaronic transition can be obtained by taking $n_d = 1/2$,
which gives $\alpha^2>1/4$. The presence of a fraction of
singly occupied states increase the critical value of $\alpha$ needed for a
bipolaronic states to be formed  (see Fig. \ref{fig:PD}).
We notice that the spinful PDF for $n_d = 1/2$  maps onto the spinless one
after the usual  rescaling ($\lambda/2 \rightarrow \lambda$  and $2\ad
\rightarrow \ad$).

Finally let us compute the phonon propagator in the anti-adiabatic regime. Using
 (\ref{eq:LangFirsovTrasf}) we obtain
\beq
\label{eq:DphLF}
D(t)=D_0(t)+4\alpha^2 s\Xi_d(t) 
\eeq
where $D_0(t)$ is the correlation function of the harmonic oscillator of
frequency $\omega_0$ and
\beq
\Xi_d(t)=-i \left< T \sum_{\sigma,\sigma^\prime} (n_\sigma(t)-\frac{1}{2})
(n_{\sigma^\prime}(0)-\frac{1}{2})\right>
\eeq
is the density-fluctuation correlation function
\cite{noteLF}.
Let us first examine a limit case in which both the spinful and spinless fermions
 are insulating, i.e., the atomic (zero hybridization) case $D=0$.
In this limit the density is a constant of motion and the Fourier transform
of (\ref{eq:DphLF}) is
we obtain in the frequency domain:
\beqa
\label{eq:DphLFatomic1}
D(\omega)=D_0(\omega)-i4\alpha^2 \frac{1}{4}\left(\frac{1}{\omega+i\delta}-\frac{1}{\omega-i\delta} \right)\\
\label{eq:DphLFatomic2}
D(\omega)=D_0(\omega)-i4\alpha^2 \left < n_\uparrow n_\downarrow \right >\left(\frac{1}{\omega+i\delta}-\frac{1}{\omega-i\delta} \right).
\eeqa
respectively in the spinless and in the spinful case.
From Eqs. (\ref{eq:DphLFatomic1}) and (\ref{eq:DphLFatomic2}) we see that the 
pole associated to the density fluctuation occurs at zero frequency.
In other words the freezing of the density
fluctuations which occurs at strong coupling imply a vanishing of a low frequency contribution
in the phonon DOS. Of course all the phonon spectral weight remains at frequency
$\pm \omega_0$.	
Upon including the hopping term, the situation changes drastically
depending on the occurrence of a MIT. If a MIT takes place,
the vanishing of the zero-frequency spectral weight is
attained at finite coupling ($\lambda_{MIT}$) on the other hand in the spinless
case this occurs only asymptotically.
To corroborate such a prediction which explains the behavior shown in Fig.
\ref{fig:data_antiad} we perform an approximate calculation of the
correlation function (\ref{eq:DphLF}), which gives for the spinless case and
in Matsubara frequencies,
$\Xi_d$ reads
\beq
 \Xi_d(i\omega_n)= \frac{1}{\beta}\sum_m G_p(i\omega_m-i\omega_n).
G_p(i\omega_m)
\eeq
If we consider the motion of the spinful electron within the CPA approximation as a
motion in a random bimodal potential \cite{Vollhardt}, then the same expression
holds also in the spinful case, and the differences are only provided by the
 form of $G_p(i\omega_m)$.

Introducing the spectral representation of $G_p$, performing the sums and the
analytical continuation to real frequencies we obtain
\beq
\label{eq:Chid}
\Xi_d(\omega)=\int_0^\omega d\nu \rho_p(\nu) \rho_p(\nu-\omega)
\eeq
where $\rho_p(\omega)=-Im G_p(\omega+i\delta)/\pi$.
Notice that if we consider the atomic limit $t=0$ we recover
(\ref{eq:DphLFatomic1}),  but not  (\ref{eq:DphLFatomic2}).
This is a drawback of the CPA approximation for the density correlations.
We adopt however Eq.  (\ref{eq:Chid}) as a CPA prescription 
which qualitatively predicts
the disappearance of the phonon spectral weight at the MIT in the
spinful case as we shall see explicitly below.

\subsection{Discussion of the results in the anti-adiabatic regime}

\begin{figure}[htbp]
\begin{center}
\includegraphics[scale=0.33,angle=270]{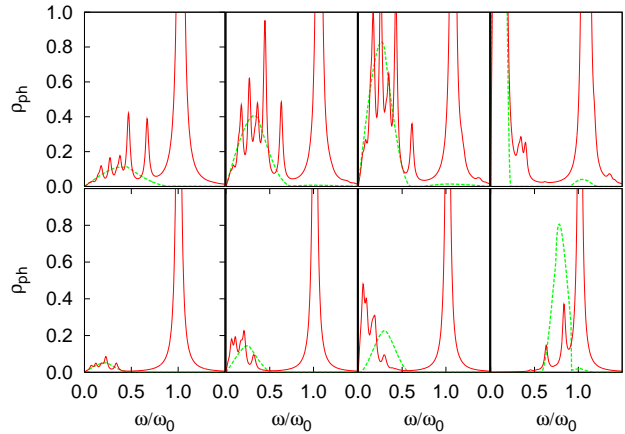}
\end{center}
\caption{(color online) Phonon DOS in the spinless (upper panel) and spinful (lower panel) cases
in the anti-adiabatic regime ($\ad=2.0$ spinless, $\ad=4.0$ spinful).
Solid curves are DMFT results, dashed lines the LF for the approximation for the charge fluctuation contribution
$\Xi_d$ in Eq. (\ref{eq:DphLF}).
The values of $\lambda$ are in the spinless (spinful case) from left to right:
$\lambda=0.4$ (0.2), $\lambda=1.2$ (0.6),$\lambda=2.0$ (1.0),$\lambda=5.2$ (2.6).
}
\label{fig:DOSPHvsLF}
\end{figure}

In Fig. \ref{fig:DOSPHvsLF} we compare the phonon DOS from DMFT-ED with the  LFCPA
described above.
A strong resonance of phononic origin is always present around $\omega_0$,
together with a low-energy broad structure due to electron density fluctuations
predicted by HLFA (\ref{eq:DphLF}).

As in the adiabatic case, but here even more clearly, we see that the  MIT implies
the disappearance of low-frequency weight in the phonon DOS.
The disappearance takes place exactly at the MIT, i.e., for a much smaller
coupling than bipolaron formation.
In a sense, the effect of the electrons on the low-energy part of the phonon
spectrum is always similar to the antiadiabatic regime.
A noticeable difference between the present antiadiabatic case and the adiabatic
regime is instead present in the high energy part of the phonon spectrum,
 which in this case is basically unperturbed in the present antiadiabatic regime.
We notice incidentally that, unlike the present case, phonon softening has
actually been found in spinless Holstein model in one dimensions\cite{Fehske,charles}.
There however a MIT is induced by Peierls dimerization and the phonon
softening is a precursor of the MIT just like in our adiabatic case where
$\lambda_s<\lambda_{MIT}$.

In our previous publication \cite{Max1} we have found a softening which is not
complete even in the spinful by estimating the renormalized phonon frequency
$\Omega$ from the $\omega_n=0$ limit of the  phonon self-energy in the imaginary  frequency
($\Omega^2=\omega^2_0+2\omega_0\Pi(i\omega_n=0))$).
We observe here that this
quantity is a good estimate for the phonon frequency only when
the majority of the phonon spectral weight is concentrated around some frequency, and
the spectrum is not divided in several features. A straightforward calculation gives
\beq
\label{eq:thermodynamicOmega}
\frac{1}{\Omega^2}= \int_0^\infty d\omega \rho_{ph}(\omega)\frac{1}{\omega},
\eeq
which implies that all the different features of the phonon propagator
contribute to $\Omega$,  therefore leading to an incomplete softening also in
the case in which the lowest phonon pole actually completely softens.
\begin{figure}[htbp]
\begin{center}
\includegraphics[scale=0.33,angle=270]{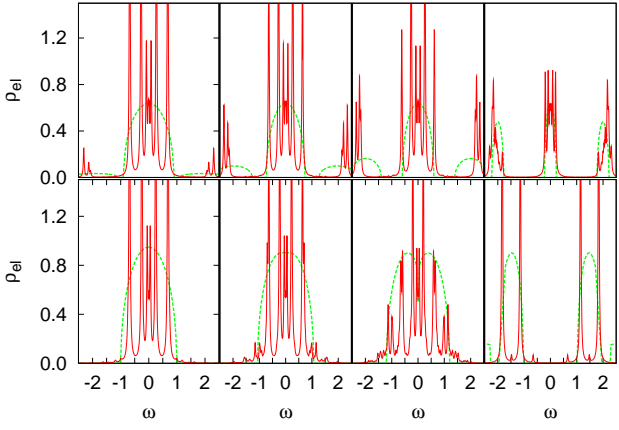}
\end{center}
\caption{(color online) DOS in the spinless (upper panel) and spinful (lower panel) cases
in the anti-adiabatic regime ($\ad=2.0$ spinless, $\ad=4.0$ spinful).
Solid curves are DMFT results, dashed lines the LF approximation.
The values of $\lambda$ are the same of Fig. \ref{fig:DOSPHvsLF}.
}
\label{fig:DOSvsLF}
\end{figure}

In Fig. \ref{fig:DOSvsLF} the electron DOS is compared with the results of
HLFA. The different behavior of spinless vs spinful cases is marked here
clearly by the presence of the MIT in the former at a value of the coupling
which is much less that that of the (bi)polaron crossover. HLFA correctly
 catches the gross behavior of the DOS. Notice that at the strong
coupling the CPA employed to obtain the lower diagrams accurately reproduces
the position of the side bands.

The different behavior of spinless vs spinful system can be easily understood
in terms of strong coupling anti-adiabatic perturbation theory for the original
lattice problem \cite{Freericks-strong}. Here we sketch an alternative
approach based on the
equivalent HIM (\ref{eq:Anderson_Holstein}) after the LF transformation
(\ref{eq:AndersonLF})
In the spinful case at second order in hybridization $V_\kvec$ the following
Kondo hamiltonian \cite{Hewson}
can be obtained \cite{Cornaglia-condmat}:
\begin{eqnarray}
\label{eq:KondoHIM}
H &=& H^{\prime}+J_{\parallel}\sum_\kvec v_k\rho^z_f \rho^z_c(\kvec)+
\nonumber\\
&+&\frac{J_{\perp}}{2}\sum_\kvec v_k (\rho^+_f \rho^-_c(\kvec) + \rho^-_f \rho^+_c(\kvec))
\end{eqnarray}
where $H^\prime$ contains all the terms of  (\ref{eq:AndersonLF}) which are not
proportional to $V_\kvec$,
$V^2=\sum_k|V_\kvec|^2$ and $|v_\kvec|^2=|V_\kvec|^2/V^2$.
$\rho$ are pseudo-spin operators corresponding to bath ($\rho_c$) or to
impurity ($\rho_f$) states,
and $J$'s are anisotropic
Kondo couplings (see also Eq. (8) of Ref.\onlinecite{Cornaglia})
\beq
J_{\parallel,\perp} =
\frac{8V^2}{D}\sum_m (\pm)^m
\frac{e^{-\alpha^2}\alpha^{2m}}{m!(m\ad+\lambda/2)},
\eeq
where the $+$ ($-$) sign is taken for the $\parallel$ ($\perp$)
coupling.
In the spinless case the processes
leading to $J_{\perp}$ do not exist while the remaining $J_{\parallel}$ is
solely associated to charge fluctuations i.e. it will correspond to a
$f$-charge $c$-charge interaction in the effective Hamiltonian
\beq
H = H^\prime+J_{\parallel}\sum_k v_k\rho^z_f \rho^z_c(\kvec)
\eeq
where now $\rho^z_f=f^\dagger f-1/2$.
No Kondo effect is expected for this Hamiltonian, while in the
spinful case the model will display a Kondo effect associated to the pseudo-spin fluctuations.
A strong coupling estimates of $J_{\parallel}$ gives
\beq
J_{\parallel}\simeq \frac{8V^2}{D\lambda}
\eeq
and $J_{\perp}/J_{\parallel} \propto \exp(-2\alpha^2)$ which means an
exponential suppression of the superconductivity versus charge
correlation at strong coupling due to retardation effects
\cite{Cornaglia-condmat}. In the anti-adiabatic
limit $\ad\rightarrow \infty$ we recover an isotropic
Kondo Hamiltonian $J_{\perp}=J_{\parallel}$.

It is worth to note that the range of validity of these expansions is somewhat
larger than the anti-adiabatic regime. Indeed the minimum energy scale of
Hamiltonian $H^\prime$ should be larger that the largest Kondo coupling
just evaluated i.e. $J_{\parallel}$ at strong coupling.
Therefore supposing $\omega_0$ to be smaller that the bandwidth of the "bath"
electrons and of polaronic energy we get $\ad >  {4V^2}/{D^2 \lambda}$.

\section{Conclusions}
\label{sec:conclusions}

We have presented a thorough study of the Holstein model comparing the
spinful and the spinless Holstein model at half-filling with the main purpose
of identifying in the most unambiguous way the polaron crossover, and
disentangling the tendency to form polarons from the 
metal-insulator transition associated to bipolaronic binding.
The role of the phonon frequency, determining whether the system
is adiabatic or antiadiabatic (slow/fast phonon dynamics) has also been
addressed with care.
We computed various observables within Dynamical Mean-Field Theory,
solving the associated impurity model by means of Exact Diagonalization.
By comparing these numerically exact results with approximate
analytical schemes suitable for the different limiting regimes we
have identified different mechanisms for the appearance of polaronic
features in the different regimes.

In the adiabatic regime a Born-Oppenheimer scheme has been devised to deal
with both small and strong coupling regime. In the anti-adiabatic case a
more standard Lang-Firsov approach for the impurity model has been applied.

While in the adiabatic case the polaron crossover and the bipolaronic MIT occur
for similar values of the coupling, in the anti-adiabatic case pairing occurs
for smaller coupling than polarization. Therefore there is a large region
of the phase diagram (whose size increases with the phonon frequency) in
which pairs are formed, but no lattice polarization is associated.
At the bipolaronic MIT the vanishing of quasi particle spectral weight at the
Fermi level \cite{Max1} is accompanied by a softening of the phonon
mode associated to low-energy charge fluctuations. The absence of such
a complete softening in the spinless case is a further signature of the
absence of insulating behavior.

Within DMFT the Mott-Hubbard transition is associated to the Kondo effect
of the Anderson impurity model. Analogously the pairing transition of
attractive models is related to the Kondo effect once a pseudospin
whose components are the s-wave superconductivity and charge ordering
are introduced.
In the Holstein model we have a mapping onto an anisotropic pseudospin Kondo model, in
which pairing correlations are greatly reduced by retardation effects but
nonetheless survives due to phonon quantum fluctuations. At the MIT the
Kondo peak associated to quasiparticle properties shrinks to zero and phononic low energy
features disappears.

An interesting feature of the polaron crossover in both spinful and spinless
case is the failure of the BO approximation in the crossover region. At the
crossover phonon and electron state becomes entangled leading to a breaking of
the BO approximation even in the adiabatic regime.
We plan to extend the
Born-Oppenheimer scheme here introduced to the more involved situation in
which the electrons also feel a Hubbard-like repulsion, a subject
which has been the subject  of different
numerical analyses, using ED  \cite{StJ,CoreaHoHu,PhS},
 and NRG techniques \cite{HewsonHoHu}.

The authors acknowledge useful discussions with E. Cappelluti and C. Castellani.
This work was supported by INFM PRA-Umbra and MIUR-Cofin 2003 matching funds
programs.

\appendix
\section{Perturbation Theory of the TSPM}
\label{app:App_pert}

\subsection{Spinless Case}

The spinless case is obtained by taking $s=1$ in  (\ref{eq:TSPM-definition}).
We  will treat the hybridization term of (\ref{eq:TSPM-definition})
as a perturbation of the free
Hamiltonian defined by $H_0=-2\e\left(f^+f-\frac{1}{2}\right)\sz -\D\sx
+\sum_k\e_k c^+_{k}c_{k}$.
We take the label $k$ to assume symmetric values with respect to zero.
We impose the particle-hole symmetry at half-filling by
means of the following conditions:
\begin{equation}\label{eq:cond}
\begin{split}
V_k &= - V_{-k} \\
\e_k&= -\e_{-k}.
\end{split}
\end{equation}
If conditions (\ref{eq:cond}) are obeyed, Eq. (\ref{eq:TSPM-definition}) is
invariant for the following transformations:
\begin{displaymath}
\begin{split}
& f^{\pm}  \to  f^{\mp}\\
& c_k^{\pm} \to  c_{-k}^\mp \\
& \sz \to -\sz \\
& \sx \to \sx
\end{split}
\end{displaymath}
where ($f^-\equiv f$, $c^-\equiv c$).
It is easily verified that the previous transformations are generated
by the unitary operator $U=\exp [(bf-f^+b^+
+\sum_k(d_{-k}c_k-c^+_{k}d^+_{-k})+i\sx)\pi/2]$ ($b$ and $d_k$ are the
new particles).
Of course the total particle number $N=f^+f +\sum_kc^+_kc_k$ is
conserved.

$H_0$ is easily studied. This Hamiltonian is diagonal in the
fermions. It is immediate to treat the band electrons term.
The spectrum and the eigenvectors of $H_0$ are given in Eqs (\ref{eq:eigvecspinless})
and (\ref{eq:eigvalspinless}).
Now we introduce
\begin{eqnarray}
\label{eq:deftilde}
|\widetilde{v^+_\s}\rangle &=&
 \frac{1}{\sqrt{2\l(\l-\s\e)}}\begin{pmatrix}
 \D \\ \e-\l\s
 \end{pmatrix}
 \nonumber\\
|\widetilde{v^-_\s}\rangle &=&
 \frac{1}{\sqrt{2\l(\l+\s\e)}}\begin{pmatrix}
 \e+\l\s \\ \D
 \end{pmatrix}.
\end{eqnarray}

The following properties hold:
\begin{eqnarray}
\label{eq:identityspinless}
\langle v_\s^+ |v_{\s'}^-\rangle &=&\langle v_{\s'}^- |v_\s^+\rangle =
\frac{\e}{\l} \delta_{\s\s'}-\s\frac{\D}{\l} \delta_{\s-\s'}
\nonumber\\
 \sum_{\s=\pm 1} |\widetilde{v^\a_\s}\rangle \langle \widetilde{v^\a_\s} | &=&
\mathbf{1}.
\end{eqnarray}
where $\a =\pm1. $
The free propagator of the impurity is defined as:
\begin{equation}
G_0(\tau) = - \frac{\text{tr} \left( e^{-\b H_0}T_\t f(\t)f^+(0) \right)}{\text{tr} \left(e^{-\b H_0}\right)}.
\end{equation}
Using (\ref{eq:identityspinless}) we get
\begin{eqnarray}
G_0(\tau) &=& -\frac{1}{2}\theta(\t) \left
(
\frac{\cosh(2\t\l-\b\l)}{\cosh{\b\l}}\frac{\e^2}{\l^2}+\frac{\D^2}{\l^2}\right)
+ 
\nonumber\\
&+&\frac{1}{2}\theta(-\t) \left
(
\frac{\cosh(2\t\l+\b\l)}{\cosh{\b\l}}\frac{\e^2}{\l^2}+\frac{\D^2}{\l^2}\right),
\end{eqnarray}
which, after a Fourier transform, becomes 
\begin{equation}
\label{eq:libero}
 G_0(i\o_n) =
\frac{1}{2}\sum_\s G_{0\s}(i\o_s),
\end{equation}
where we have introduced $G_{0\s}$ given by
\begin{equation}
G_{0\s}(i\o_s) =\frac{1}{2}
\left( \frac{\e^2}{\l^2}\frac{1}{i\o_n+2\l\s} +
\frac{\D^2}{\l^2}\frac{1}{i\o_n}\right)
\end{equation}
and $\o_n=(2n+1)\pi/\b$.

The hamiltonian (\ref{eq:TSPM-definition}) is quadratic in the fermions
but the spin
operators generate an effective interaction. For this reason
we can not expect a finite set of equations of motion if the band
contains ``many'' electrons.

We introduce the quantity :
\begin{equation}\label{eq:not}
G_{ff^+}(\t) = - \frac{1}{Z}\text{tr} \left( e^{-\b H}T_\t f(\t)f^+(0) \right),
\end{equation}
we find:
\begin{equation}\label{eq:eom}
\dot{G}_{ff^+}(\t)=-\delta(\t)+2\e G_{\sz ff^+}(\t) -\sum_k
G_{c_kf^+}(\t),
\end{equation}
where adopting the same notations as for Eq. (\ref{eq:not}), $G_{\sz
ff^+}(\t)$ is defined as
\begin{displaymath}
G_{\sz ff^+}(\t) = - \frac{1}{Z}\text{tr} \left( e^{-\b H}T_\t
\sz(\t)f(\t)f^+(0) \right).
\end{displaymath}
$G_{c_kf^+}$ is defined similarly. The equations of motion for the r.h.s. terms of
Eq. (\ref{eq:eom}) are
\begin{displaymath}
\begin{split}
& \dot{G}_{c_kf^+}(\t) = -\e_kG_{c_kf^+}(\t) -V_kG_{ff^+}(\t); \\
& \dot{G}_{\sz ff^+}(\t) = i\D G_{\sy ff^+}(\t) +2\e
G_{ff^+}(\t)-\\
&-\sum_k V_kG_{\sz c_kf^+}(\t).
\end{split}
\end{displaymath}
The first equation does not give rise to any new quantity, but the second
involves $G_{\sy ff^+}$ and $G_{\sz c_kf^+}$. Therefore, we should
derive the equations of motion for
these terms,  and proceed until the iterations do not generate  any new quantity.
Unfortunately this procedure does not seem to be convergent. Terms like $G_{\sy
c_kf^+}$, $G_{\sx ff^+}$, etc. are generated and the equation for
$G_{\sy c_kf^+}$ introduces the correlator
\begin{widetext}
\begin{displaymath}
G_{D\sx c_k f^+}(\t)= - \frac{1}{Z}\text{tr} \left\{ e^{-\b H}T_\t
\left [ -2\e\left(f^+(\t)f(\t)
-\frac{1}{2}\right)\sx(\t)c_k(\t)f^+(0)\right] \right\}.
\end{displaymath}
\end{widetext}
The equation of motion for the latter involves, among others,
\begin{displaymath}
- \frac{1}{Z}\text{tr} \left\{ e^{-\b H}T_\t
 2\e \sum_p V_p c^+_p(\t)f(\t)\sx(\t)c_k(\t)f^+(0) \right\}.
\end{displaymath}
It is evident how a hierarchy of correlators with increasing complexity is thus generated by this
process.
The impossibility of further pursuing this approach forces us to resort to perturbation theory.

We aim to compute the interacting propagator:
\begin{displaymath}
G(\tau) = - \frac{\text{tr} \left( e^{-\b H}T_\t f(\t)f^+(0) \right)}{\text{tr} \left(e^{-\b H}\right)}.
\end{displaymath}
We resort to second
order perturbation theory in the $V_k$ coefficients. If $Z_0=\text{tr}
(e^{-\b H_0})$, $V$ is the hybridization and, as usual, $V(\t)=e^{\t
H_0}V e^{-\t H_0}$, we have:
\begin{widetext}
\begin{displaymath}
G^{(2)}(\tau) = -\frac{1}{2 Z_0 }\text{tr} \left( e^{-\b
H_0}T_\t f(\t)\int_0^\b d\t_1d\t_2 V(\t_1)V(\t_2)f^+(0)
\right)_{\text{connected}}.
\end{displaymath}
Taking the Fourier transform
and using again Eq. (\ref{eq:identityspinless}), we obtain the second-order perturbative
result
\begin{equation}\label{eq:sec}
\begin{split}
G^{(2)}(i\o_n) &= \frac{1}{2}\sum_\s  \left( \frac{\e^2}{\l^2}\frac{1}{i\o_n+2\l\s} +
\frac{\D^2}{\l^2}\frac{1}{i\o_n}\right)^2\sum_k\frac{V_k^2}{i\o_n-\e_k}
+\\
& +\sum_\s \left[\frac{\e\D}{\l^2}
\left(\frac{1}{i\o_n+2\l\s}-\frac{1}{i\o_n}\right) \right]^2
\frac{1}{\b}\sum_{k,p_n}\frac{V^2_k}{ip_n-\e_k}\cdot
\frac{\s\tanh(\b\l)}{i\o_n-ip_n+2\l\s},
\end{split}
\end{equation}
\end{widetext}
where $\o_n$ and $p_n$ are fermionic momenta. The interpretation of the
first term of Eq. (\ref{eq:sec}) is straightforward since in can be
written as
\begin{equation}
G^{(2)}_A(i\o_n)=\frac{1}{2}\sum_\s
G_{0\s}(i\o_n)\Sigma_{\text{ibr}}(i\o_n)G_{0\s}(i\o_n),
\end{equation}
where we have introduced the hybridization self energy
$\Sigma_{\text{ibr}}(i\o_n)
=\sum_k\frac{V_k^2}{i\o_n-\e_k}$. $G_{0\s}$, defined in the last line of
Eq. (\ref{eq:libero}), can be interpreted as the free propagator of the
impurity for a given spin $\s$. As a first approximation for the
interacting propagator we can sum the Dyson series for the $A$-terms,
obtaining the BOCPA approximation (\ref{eq:GBOCPA}) discussed
in Sec. \ref{sec:adiabatic}.

The second term (henceforth called $B$-term) of Eq. (\ref{eq:sec}),
\begin{widetext}
\begin{equation}\label{eq:GB}
G^{(2)}_B(i\o_n)=\sum_\s \left[\frac{\e\D}{\l^2}
\left(\frac{1}{i\o_n+2\l\s}-\frac{1}{i\o_n}\right) \right]^2
\frac{1}{\b}\sum_{k,p_n}\frac{V^2_k}{ip_n-\e_k}\cdot
\frac{\s\tanh(\b\l)}{i\o_n-ip_n+2\l\s},
\end{equation}
\end{widetext}
is surprising for the following reasons:

{\it i)} it does not contain the free propagator;

{\it ii)} the two external legs, namely the two factors
$\left[\frac{\e\D}{\l^2}
\left(\frac{1}{i\o_n+2\l\s}-\frac{1}{i\o_n}\right) \right]$ carry a
fermionic frequency $i\o_n$ but the sum of the residues of the poles
is zero  so this object does not carry charge;

{\it iii)} the internal
line $\frac{\s\tanh(\b\l)}{i\o_n-ip_n+2\l\s}$ carries a bosonic
frequency and, when summed over $\s$, again has residues adding up to
zero.

The last two properties suggest us to interpret the propagators
of $B$-type $G_B^{(2)}$ as  spin correlators.

To put this interpretation on more solid ground, we consider the spin correlator defined as:
\begin{equation}
\label{eq:spin_corr}
S^{zx}_\a(\t) =- \frac{\text{tr}_\a \left( e^{-\b H_0}T_\t
\sz(\t)\sx(0) \right)}{\text{tr}_\a \left(e^{-\b H_0}\right)}
\end{equation}
which is explicitely
\begin{equation}
S^{zx}_\a(\t) =-\frac{\sum_{\s}\langle v^\a_\s| e^{-\b H_0}T_\t
\sz(\t)\sx(0) | v^\a_\s \rangle}{\sum_{\s}\langle v^\a_\s| e^{-\b
H_0} | v^\a_\s \rangle}.
\end{equation}
The imaginary time ordering $T_\t$ is defined taking into account the
canonical anticommutation relation: $\{\sz,\sx\}=0$:
\begin{eqnarray*}
T_\t \sz(\t)\sx(\t') &=& \theta(\t-\t')\sz(\t)\sx(t')-\\
&-&\theta(\t'-\t)\sx(t')\sz(\t).
\end{eqnarray*}
The $\t\to \t'$ ordering limit can be defined arbitrarily according to
convenience. It is easy to find that $S^{zx}_\a(\t)$ is antiperiodic
in $\t$: if $\t<0$ but $\t+\b>0$ then
\begin{displaymath}
S^{zx}_\a(\t+\b)=-S^{zx}_\a(\t).
\end{displaymath}
Of course only odd frequencies ($\o_n
= \frac{\pi}{\b}(2n+1)$) contribute to the Fourier series
of $S^{zx}_\a(\t)$:
\begin{equation}\label{eq:corr_spin}
S^{zx}_\a(i\o_n)
=\a\left[\frac{\e\D}{\l^2}
\left(\frac{1}{i\o_n+2\l\s}-\frac{1}{i\o_n}\right) \right].
\end{equation}
Therefore, the spin-correlator reproduces the first factor in Eq. (\ref{eq:GB}).
It is worth noticing that if we consider the correlator:
\begin{displaymath}
S^{xz}_\a(\t) =- \frac{\text{tr}_\a \left( e^{-\b H_0}T_\t
\sx(\t)\sz(0) \right)}{\text{tr}_\a \left(e^{-\b H_0}\right)},
\end{displaymath}
we obtain the same result (\ref{eq:corr_spin}) (the equal-time ordering
of $\sx$ and $\sz$ is irrelevant).

We still have to explain the last factor of
Eq. (\ref{eq:GB}). Now the product $\sz\sx$ is proportional to $\sy$,
so it is natural to consider the correlator:
\begin{equation}\label{eq:yy}
S^{yy}_\a(\t) =- \frac{\text{tr}_\a \left( e^{-\b H_0}T_\t
\sy(\t)\sy(0) \right)}{\text{tr}_\a \left(e^{-\b H_0}\right)}.
\end{equation}

It is not obvious how to define a time-ordering operator in this case.
Since $\sy^2=1$, a fermionic time ordering does not seem
appropriate ($f^2=0$ for a fermion). For this reason we define a time
ordering taking into account the
canonical commutation relations: $[\sy,\sy]=0$:
\begin{displaymath}
T_\t \sy(\t)\sy(\t') = \theta(\t-\t')\sy(\t)\sy(t')
+\theta(\t'-\t)\sy(t')\sy(\t).
\end{displaymath}
It can be observed that $S^{yy}_\a$ is actually independent on $\a$,
so we will omit this index.
For the Fourier transform we find:
\begin{equation}
S^{yy}(i\o_n) =-\sum_\s\frac{\s\tanh(\b\l)}{i\o_n+2\l\s},
\end{equation}
with bosonic frequencies ($\o_n=\frac{2\pi}{\b}n$).
This is exactly the missing piece of information to complete the
interpretation of Eq. (\ref{eq:GB}).

Two useful results are
\begin{displaymath}
\begin{split}
&S^{zz}(i\o_n)
=-\frac{\D^2}{\l^2}\sum_\s\frac{\s\tanh(\b\l)}{i\o_n+2\l\s}\\
&S^{xx}(i\o_n)
=-\frac{\e^2}{\l^2}\sum_\s\frac{\s\tanh(\b\l)}{i\o_n+2\l\s},
\end{split}
\end{displaymath}
where $S^{xx}$ and $S^{zz}$ are defined analogously to $S^{yy}$ and
the frequency $\o_n$ is again bosonic.

In order to define the diagrammatic expansion used in Fig.
\ref{fig:pertVk}, we need four types of lines (propagators)
and two vertices. The lines are:
\begin{displaymath}
\begin{split}
\text{impurity: } \langle f^+ f \rangle_\s = \frac{\e^2}{\l^2}\frac{1}{i\o_n+2\l\s} +
\frac{\D^2}{\l^2}\frac{1}{i\o_n} &= \parbox{60mm}
{\includegraphics[scale=0.3,angle=0]{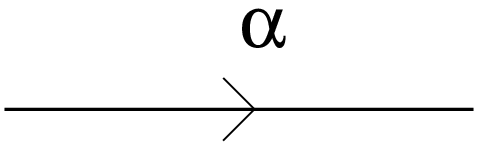}} \\
& \\
\text{bath: }\langle c^+_k c_k \rangle =
\frac{1}{i\o_n-\e_k} &= \parbox{60mm}
{\includegraphics[scale=0.3,angle=0]{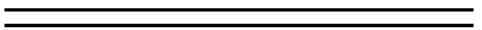}}\\
& \\
\text{$f-$spinon: }\langle \sz \sx \rangle = \frac{\e\D}{\l^2}
\left(\frac{1}{i\o_n+2\l\s}-\frac{1}{i\o_n}\right)
&=\parbox{60mm}{\includegraphics[scale=0.3,angle=0]{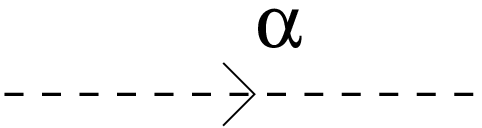}}\\
& \\
\text{$b-$spinon: }\langle \sy \sy \rangle =
\frac{\s\tanh(\b\l)}{i\o_n+2\l\s}&=\parbox{60mm}
{\includegraphics[scale=0.3,angle=0]{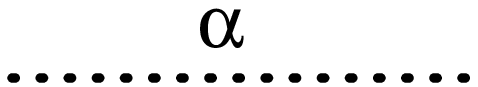}}\\
\end{split}
\end{displaymath}
The vertices are:
\vskip 0.5cm
\parbox{30mm}{\includegraphics[scale=0.3,angle=0]{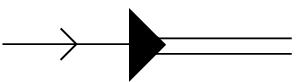}}
\parbox{30mm}{\includegraphics[scale=0.3,angle=0]{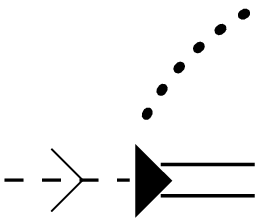}}
\vskip 0.5cm
where each bold triangle denotes the hybridization constant $V_k$.

Let us detail the Feynman rules (derived from the 2nd order
result). We start from the conservation laws.  Each line carries a
spin and a fermion number. The fermion number $n_f$ is defined to be
$1$ for $c_k$, $f$ and f-spinon lines and $0$ for the b-spinon
line. The spin number $n_\s$ is $1$ for the spinon lines and equals
$0$ for the $f$ and $c_k$ lines.

{\it{(a)}} Fermion and spin numbers are separately conserved in every
of Eqs. (\ref{eq:DphLFatomic1},\ref{eq:DphLFatomic2}). 
  allowed process.
For the purpose of computing $G(i\o_n)$, the band
electrons and the b-spinons act like virtual particles.
While this observation was relatively expected, a much more
 striking result of Eq. (\ref{eq:sec}) is that
the f-spinon lines can also be real:

{\it{(b)}} The diagrams contributing to $G(i\o_n)$ are obtained
drawing self energy diagrams with
external lines
\vskip 0.5cm
\parbox{20mm}{\includegraphics[scale=0.3,angle=0]{fig18.eps}}
\parbox{20mm}{\includegraphics[scale=0.3,angle=0]{fig19.eps}}
\vskip 0.5cm
According to this rule the f-spinon line should contribute to zero
order, whereas it does not. The reason is that a sum over the $\a$
index is actually implicit and, according to Eq. (\ref{eq:corr_spin}), this sum
vanishes.
As a consequence of {\it{(a)}}, self-energy diagrams with external
legs of different types are forbidden.

{\it{(c)}} All the lines contributing to a given diagram have the
same $\s$ label (the $\s$ label should not be confused with
the $n_\s$ number). We have to sum over
$\s$, with a coefficient $1/2$ if the two external legs are (free)
impurity propagators, and $1$ if they are f-spinons. (One may try to
justify this rule taking into account the sum over the $\a$ index: the
impurity selects a value for this index, depending on the time
ordering, while this is not the case for the spin operators, according to
Eqs. (\ref{eq:spin_corr}) and (\ref{eq:corr_spin})). These rules lead to diagrammatic
representation of the perturbation expansion shown in Fig. \ref{fig:pertVk}.

\subsection{Spinful case}

The eigenstates and eigenvalues for $H_0
=-\e\sum_{\s=\pm}\left(f_\s^+f_\s-\frac{1}{2}\right)\sz -\D\sx$ are easily
found. Denoting with $|v\rangle$ the singly occupied states and with $|u\rangle$ the doubly
occupied or empty states we get eigenvectors and eigenvalues reported in Eqs..
(\ref{eq:eigvecspinful1}),(\ref{eq:eigvecspinful2}) and (\ref{eq:eigvalspinful}).

Using the same notation of Eq. (\ref{eq:deftilde}) we get
\begin{eqnarray}
\label{eq:sca}
\langle \widetilde{u_{\s'}^\a} |\widetilde{v_{\s}^+}\rangle &=&
\frac{\D+\a\s(\l-\s\e)}{2\sqrt{\l(\l-\s\e)}}
\nonumber \\
\langle \widetilde{u_{\s'}^\a} | \widetilde{v_{\s}^-}\rangle &=&
-\a\frac{\D-\a\s(\l+\s\e)}{2\sqrt{\l(\l+\s\e)}}
=-\a\langle \widetilde{u_{\s'}^\a} | \widetilde{v_{-\s}^+}\rangle,
\end{eqnarray}
where $\a,\s=\pm 1$. The scalar products (\ref{eq:sca}) do not depend
on $\s'$.  The
completeness relations are:
\begin{eqnarray}
\sum_{\s=\pm 1} |\widetilde{v^\a_\s}\rangle \langle \widetilde{v^\a_\s} | =
\mathbf{1};
\nonumber \\
 \sum_{\a=\pm 1} |\widetilde{u^\a_\s}\rangle \langle \widetilde{u^\a_\s} | =
\mathbf{1}.
\end{eqnarray}

Let us define the following propagators:
\begin{eqnarray}
\label{eq:g12}
G_{1\s}&=&\frac{1}{2\l}\left(
\frac{\l-\D}{i\o_n+\s(\l+\D)}+\frac{\l+\D}{i\o_n+\s(\l-\D)}\right);
\nonumber \\
G_{2\s}&=&\frac{1}{2\l}\left(
\frac{\l-\D}{i\o_n+\s(\l+\D)}+\frac{\l+\D}{i\o_n+\s(-\l+\D)}\right).
\end{eqnarray}
The free propagator is given by:
\begin{equation}
\label{eq:free}
G_0(i\o_n)=\frac{1}{Z_+}\sum_{\s}\left( \cosh(\b\l)G_{1\s} +
\cosh(\b\D)G_{2\s}\right).
\end{equation}
where $Z_+=2\cosh(\b\l) + 2\cosh(\b\D)$.
The second-order perturbation theory in $V_k$  displays a
structure similar to the spinless case. After a straightforward
calculation we find, including zero order term:
\begin{widetext}
\begin{equation}
\label{eq:sec_2}
\begin{split}
G^{(2)}(i\o_n) = \frac{1}{Z_+}\sum_\s  \Big(
&\cosh(\b\l)\big[G_{1\s}(i\o_n)+G_{1\s}(i\o_n)\Sigma_{\text{ibr}}(i\o_n)G_{1\s}(i\o_n)\big]+
\\
+&\cosh(\b\D)\big[G_{2\s}(i\o_n)+G_{2\s}(i\o_n)\Sigma_{\text{ibr}}(i\o_n)G_{2\s}(i\o_n)\big]\Big)\\
+\frac{1}{Z_+}\sum_\s \frac{1}{\b} &
\sum_{k,p_n}\frac{V^2_k}{ip_n-\e_k} \frac{\e^2}{\l^2}\Big\{\frac{\s
\sinh(\b\l)}{i(\o_n-p_n)+2\l\s} \big(
\frac{1}{i\o_n+\s(\l+\D)}-\frac{1}{i\o_n+\s(\l-\D)}\big)^2 +\\
& \quad + \frac{\s
\sinh(\b\D)}{i(\o_n-p_n)+2\D\s} \big(
\frac{1}{i\o_n+\s(\l+\D)}-\frac{1}{i\o_n+\s(-\l+\D)}\big)^2\Big\}.
\end{split}
\end{equation}
\end{widetext}
As for Eq. (\ref{eq:sec}), the two terms are of diffent type. In
particular the free propagator does not appear in the second term, the
external lines do not carry charge, and inside the loop there is a
chargeless bosonic line. As for the spinless case, let us consider the
Dyson series for the first type terms (coherent potential
approximation):
\begin{widetext}
\begin{equation}
G_{c}(i\o_n) = \frac{1}{Z_+}\sum_\s  \Big(
\frac{\cosh(\b\l)}{G_{1\s}(i\o_n)^{-1}-\Sigma_{\text{ibr}}(i\o_n)}
+ \frac{\cosh(\b\D)}{G_{2\s}(i\o_n)^{-1}-\Sigma_{\text{ibr}}(i\o_n)}
\Big).
\end{equation}
\end{widetext}
Since $\l>\D$, in the zero temperature limit $\b\to \infty$, we get
for this approximation of the propagator:
\begin{equation}
G_{c}(i\o_n)=\frac{1}{2}\sum_{\s}\frac{1}{G_{1\s}(i\o_n)^{-1}-\Sigma_{\text{ibr}}(i\o_n)}
\end{equation}
which is the BOCPA in the spinful case.


\begin{thebibliography}{99}
\bibitem{mgb2}
Y. Kong, O.V. Dolgov, O. Jepsen, and O. K. Andersen,
Phys. Rev. B {\bf 64}, 020501(R) (2001).\\
K.-P. Bohnen, R. Heid, and B. Renker,
Phys. Rev. Lett. {\bf 86}, 5771 (2001).\\
A. Y. Liu, I. I. Mazin, and J. Kortus,
Phys. Rev. Lett {\bf 87}, 087005 (2001)

\bibitem{revgunnarson}  O. Gunnarsson, Rev. Mod. Phys. {\bf 69}, 575 (1997).


\bibitem{graphite} Matteo Calandra and  Francesco Mauri, cond-mat/0506082


\bibitem{cuprates-lanzara-dastuto} A. Lanzara, P. V. Bogdanov, X. J. Zhou,
S. A. Kellar, D. L. Feng, E. D. Lu, T. Yoshida, H. Eisaki, A. Fujimori, K. Kishio,
J. -I. Shimoyama, T. Noda, S. Uchida, Z. Hussain, and Z.-X. Shen,
Nature {\bf 412}, 510 (2001);

M. d'Astuto, P. K. Mang, P. Giura, A. Shukla,
P. Ghigna, A. Mirone, M. Braden, M. Greven, M. Krisch, and F. Sette,
Phys. Rev. Lett. {\bf 88}, 167002 (2002).


\bibitem{lanzara-isot}  G.-H. Gweon, T. Sasagawa, S. Y. Zhou, J. Graf, H. Takagi, D.-H. Lee, and A. Lanzara,
 Nature, {\bf 430}, 187 (2004)


\bibitem{calvani-rev} P. Calvani, La Rivista del Nuovo Cimento {\bf 24} 1 (2001).


\bibitem{manganites}
A.J.Millis, R. Mueller and B. I. Shraiman, Phys. Rev. B {\bf 54} ,5405,
(1996).\\
P. G. Radaelli and G. Iannone, M. Marezio, H. Y. Hwang and S-W. Cheong,
J. D. Jorgensen and D. N. Argyriou , Phys. Rev. B, {\bf 56}, 8265 (1997).


\bibitem{IR-Manga}
A. Congeduti, P. Postorino, P. Dore, A. Nucara, S. Lupi,
S. Mercone, P. Calvani, A. Kumar, and D. D. Sarma, Phys. Rev. B \textbf{63},
184410 (2001).\\
P. Postorino, A. Congeduti, P. Dore, A. Sacchetti, F. Gorelli, 
L. Ulivi, A. Kumar, and D. D. Sarma, Phys. Rev. Lett.
\textbf{91} 175501 (2003).


\bibitem{oxides} A. Jayaraman, D. B. McWhan, J. P. Remeika, and P. D. Dernier
Phys. Rev. B {\bf 2}, 3751-3756 (1970).


\bibitem{organics}
E. M. Conwell and S. V. Rakhmanova Proc. Natnl. Acad. Sci., {\bf{97}}, 4556 (2000)\\
E. M. Conwell Proc. Natnl. Acad. Sci.,  {\bf{102}}, 8795 (2005)


\bibitem{Morpurgo} R. W. I. de Boer, M. E. Gershenson, A. F. Morpurgo, and V. Podzorov
Phys. Stat. Sol. A {\bf 201}, 1302 (2004)


\bibitem{depolarone} S.Ciuchi, F.de Pasquale, S. Fratini, and D.Feinberg,
Phys. Rev. B {\bf 56}, 4494 (1997).


\bibitem{frat2} S. Fratini and  S. Ciuchi, Phys. Rev. Lett. {\bf 91}, 256403 (2003).


\bibitem{Lang-Firsov} I. G. Lang  and Yu. A. Firsov, Sov.Phys. JETP {\bf
16},1301 (1963)


\bibitem{DMFTreview} A. Georges, G. Kotliar, W. Krauth, and M.J. Rozenberg, Rev. Mod. Phys. {\bf 68} 13 (1996).


\bibitem{Max1} M. Capone and S. Ciuchi, Phys. Rev. Lett. {\bf 91}, 186405 (2003)


\bibitem{Millis-adiab}
A.J.Millis, R. Mueller and B. I. Shraiman, Phys. Rev. B {\bf 54} ,5389,
(1996).


\bibitem{cdw-adiab} S. Ciuchi and F. de Pasquale Phys. Rev. B {\bf 59}, 5431 (1999).


\bibitem{NoteKF} Notice that the Holstein model in the adiabatic approximation
and at zero temperature is equivalent to a Falicov-Kimball model as was noted
first in Ref. \onlinecite{FJS}.


\bibitem{FJS} J.K.Freericks, M.Jarrell, and D.J.Scalapino, Phys. Rev. B {\bf 48},
6302 (1993).


\bibitem{NoteKF2} This expression is equivalent to the CPA free energy of the
Kimball Falikov model \cite{BrandtMielsch,ChungFreericks}


\bibitem{BrandtMielsch} U. Brandt and C. Mielsch, Z. Phys. B {\bf 75}, 365 (1989);
{\bf 79} 295 (1990); U. Brandt, A. Fledderijorann, and G. H\"ulsembeck, {\it
ibid} {\bf 81} 409 (1990); U. Brandt and  C. Mielsch {\it ibid} {\bf 82}, 37 (1991)


\bibitem{ChungFreericks} W. Chung and J. K. Freericks, Phys. Rev. B {\bf 57},
11955 (1998).


\bibitem{pata} P. Benedetti and  R. Zeyher, Phys. Rev. B {\bf 58}, 14320 (1998).


\bibitem{Engelsberg} S. Engelsberg and J.R. Schrieffer, Phys. Rev. {\bf 131}, 993
(1963).


\bibitem{noteMITCPA} Notice that in the language of TSPM the MIT
occurs at a given $\e_s(\D)$ which means by using Eqs. (\ref{eq:TSPM-parameters}) a given
$\lambda(\ad )$.


\bibitem{russianBO} E.G. Brovman and Yu. Kagan,
Sov. Phys. JETP {\bf 25}, 365 (1967)


\bibitem{storia} M. Capone, W. Stephan, and M. Grilli, Phys. Rev. B {\bf 56}, 4484 (1997);
M. Capone, C. Grimaldi, and S. Ciuchi, Europhys. Lett. {\bf 42}, 523 (1998).


\bibitem{Ranninger_spectral}
J. Ranninger, Phys. Rev. B {\bf 48}, R13166 (1993)


\bibitem{Alexandrov-Ranninger} A.
S. Alexandrov and J. Ranninger, Phys. Rev. B {\bf 45}, R13109 (1992); \\ A.
S. Alexandrov and J. Ranninger, Physica C {\bf 198}, 360 (1992).


\bibitem{Uc1Uc2neg}
M. Capone, C. Castellani, and M. Grilli, Phys. Rev. Lett. {\bf 88}, 126403
(2002); A. Toschi, P. Barone, M. Capone, and C. Castellani,
New Jour. of Phys. {\bf 7}, 7 (2005)



\bibitem{Uc1Uc2}
R. Bulla, Phys. Rev. Lett. {\bf 83},136 (1999)


\bibitem{noteLF} The averages are performed on the ground state of the
transformed Hamiltonian (\ref{eq:AndersonLF}) after HLFA approximation.


\bibitem{Vollhardt} The relation between DMFT and CPA is discussed, e.g.,  in
D.Vollhardt in {\it Correlated Electron Systems} ed. by
V.J. Emery  (World Scientific, Singapore, 1992)


\bibitem{Fehske} S. Sykora, A. H\"ubsch, K. W. Becker, G. Wellein,
and H. Fehske, Phys. Rev B {\bf 71} 045112 (2005)


\bibitem{charles} C.E. Creffield, G. Sangiovanni, and M. Capone,
Eur. Phys. J. B {\bf 44}, 175 (2005)


\bibitem{Freericks-strong} J. K. Freericks,
Phys. Rev. B {\bf 48}, 3881 (1993).


\bibitem{Hewson} A.C. Hewson {\it The Kondo Problem to Heavy Fermions}
Cambridge University press (1993).


\bibitem{Cornaglia-condmat} P.S. Cornaglia, D.R. Grempel, and H. Ness,
Phys. Rev. B {\bf 71}, 075320 (2005)


\bibitem{Cornaglia} P.S. Cornaglia, H. Ness, and D.R. Grempel
Phys. Rev. Lett. {\bf 93}, 147201 (2004)


\bibitem{StJ} G. Sangiovanni, M. Capone, C. Castellani, and M. Grilli
Phys. Rev. Lett. {\bf 94}, 026401 (2005).


\bibitem{CoreaHoHu}  G.S. Jeon, T.-H. Park, J. H. Han, H. C.
Lee, and H.-Y. Choi, Phys. Rev. B {\bf 70}, 125114 (2004)


\bibitem{PhS} M. Capone, G. Sangiovanni, C. Castellani, C. Di Castro, and M. Grilli,
Phys. Rev. Lett. {\bf 92}, 106401 (2004).



\bibitem{HewsonHoHu} W. Koller, D. Meyer, Y. Ono, A. C. Hewson, Europhys. Lett. {\bf 66}, 559 (2004);
W. Koller, D. Meyer, and A. C. Hewson, Phys.
Rev. B {\bf 70}, 155103 (2004).


\bibitem{Bulla}  D. Meyer, A.C. Hewson, and R. Bulla, Phys. Rev. Lett. {\bf 89}, 196401 (2002).


\end{thebibliography}
\end{document}